\documentstyle[aas2pp4]{article}
\newcommand{\ltwid}{\mathrel{\raise.3ex\hbox{$<$\kern-.75em\lower1ex\hbox{$\sim$}}}}
\newcommand{\bd}{\begin{description}}
\newcommand{\ed}{\end{description}}
\def\s{\scriptscriptstyle}
\begin{document}
\title{Expectations from a Microlensing Search for Planets}
\author{S.J. Peale}
\affil{Dept. of Physics\\
University of California\\
Santa Barbara, CA 93106\\
peale@io.ucsb.edu}
\begin{abstract}
The statistical distribution of the masses of planets about stars between the
Sun and the center of the galaxy is constrained to within a factor of
three by an intensive search for planets
during microlensing events. Projected separations in terms of the
lens Einstein ring radius yield a rough estimate of the distribution
of planetary semimajor axes with planetary mass.  The search consists
of following ongoing stellar microlensing events involving sources in the
center of the galaxy lensed by intervening stars with high time resolution,
1\% photometry in two colors in an attempt to catch any short time
scale  planetary
perturbations of the otherwise smooth light curve. It is assumed that
3000 events are followed over an 8 year period, but with half of
the lenses, those that are members of binary systems, devoid of planets. The
remaining 1500 lenses have solar-system-like distributions of 4 or
5 planets. The expectations from the microlensing search are extremely
assumption dependent with 56, 138, and 81 planets being detected for
three sets of assumptions involving how the planetary masses and
separations vary with lens mass. The events can be covered from 54\%
to 62\% of the time on average by high time resolution photometry from
a system of three or four dedicated two meter telescopes distributed in
longitude, so 38\% to 46\% of the detectable small mass planets (very short
perturbations of the light curve) will be missed. But
perturbations comparable to a day in length means all of the
detectable Jupiters and Saturns will in fact be detected as well as
a large fraction of the Uranuses.    The ground based
observational technique is robust, and meaningful statistics on
planetary masses and separations can be inferred from such an
intensive search, although these statistics, like the inferred data
set, will also be dependent on the assumptions about the nature of the
set of planetary systems.  Finding most of many giant and sub
giant planets outside the Einstein ring radii of their respective
stars may be a better indicator of the frequency of Earth mass planets
than direct detection of a few of the latter.  
\end{abstract}
\section{Introduction}
The nearly circular and coplanar orbits of the planets around the Sun
point to the formation of these planets from a highly dissipative
disk of gas and dust.  The inevitability of the natural formation of such a
disk during the gravitational collapse of a rotating molecular cloud
to form a star and the observational confirmation that essentially all
recently formed stars possess such a disk ({\it e.g.} Strom {\it et
al.} 1995) have led most to believe that planetary systems are a
common result of the star formation process and therefore must be
ubiquitous in the galaxy.  The recent discovery of planetary sized
bodies around several nearby stars (Mayor and Queloz, 1995; Marcy and
Butler, 1996; Butler and Marcy, 1996; Gatewood, 1996) and about the
pulsar PSR 1257+12 (Wolszczan and Frail, 1992) reinforces this
confidence.  However, except for the system around the M star Lalande
21185 (Gatewood, 1996), the inferred planets are in systems very much
unlike our own, since planets comparable to or exceeding Jupiter's mass
orbit very close to their primary stars.  In fact, those companions in
eccentric orbits about the ordinary stars are inferred to not be
planets at all but were formed
like binaries by fragmentation of a collapsing molecular cloud (Boss,
1996; Mazeh {\it et al.} 1996). This leaves massive planets comparable
to or exceeding Jupiter's mass in nearly circular orbits about 51 Peg
($m\sin{i}\sim 0.45m_{\s J}$, 4.2 day period), 55 Cnc ($m\sin{i}\sim  
0.78m_{\s J}$, 14.7 day period) and 47 UMa ($m\sin{i}\sim 2.4m_{\s J}$, 
1090 day period) contrary to our expectations of the stellar distances
at which giant planets would be formed. (However, there is a mechanism
by which such planets could migrate toward their stars after formation
provided a nebula with significant mass persists sufficiently long
(Lin, {\it et al.} 1996; Ward and Hourigan, 1989; Ward, 1996)). 
These discoveries imply  that
other planetary systems are likely to differ greatly from our own,
with totally different distributions of planetary mass and character
with distance from the star.  In particular the number of terrestrial
type planets in any system may be considerably different from
expectations based on our own solar system.   

The theory of the formation and evolution of planetary systems depends
on only one example, whereas the recent discoveries of other planets
indicate that a variety of scenarios leading to systems strikingly
different from our own may be common.  To understand the frequency
of the particular chain of events that led to a planetary
system like ours, where life was able to develop on one terrestrial
planet, we need statistics not only on the occurrence of other
planetary systems, but also on the planetary mass distributions as a
function of the planet separation from the parent star.  Giant planets
close to their primaries may preclude the existence of terrestrial
type planets in those particular systems.  The recent
discoveries of planets around other stars depended on detecting motion of
the star relative to a system center of mass with the period of the
planetary orbit. High precision Doppler spectroscopy and, in the case
of Lalande 21185, high precision astrometry were used to determine the
stellar motions.   These techniques can provide information on
planetary mass distributions and separations, where the Doppler
technique is most sensitive to planets that are close to their stars
and astrometry most sensitive to planets that are far away.
Unfortunately, these observations are very demanding so relatively few
stars can be monitored, and the observations must be continued for at least
a planetary period for secure detection.  Several decades of dedicated
effort would be required to compile adequate statistics, and neither
method is sensitive to terrestrial-mass planets at any separation.

The recently announced intention of the National Aeronautics and Space
Administration to search for planets about other stars (Beichman, 
1996) has as its goal the detection and characterization of
terrestrial type planets orbiting stars within about 10 pc from the Sun.
The instrumentation to be developed in this search depends on the
distribution of planetary masses and in particular on the frequency of
occurrence of Earth-mass planets.  Planning this program requires
the planetary statistics on a relatively short time scale, and it requires
information on the frequency of occurrence of terrestrial-mass planets
that is not obtainable by the radial velocity and astrometry
techniques.   Mao and Paczy\'nski
(1991) pointed out that a planet could be detected as a companion of a
star by perturbing the light amplification curve of a more distant
star (source) that is being gravitationally lensed (microlensing) 
by the nearer star
(lens). Later Gould and Loeb (1992) derived a probability of nearly
20\% for detecting a Jupiter about a solar mass star given that a
microlensing event was taking place by assuming that a 5\% perturbation
of the light curve was detectable.  Microlensing is a rare event, so
millions of stars must be monitored in order to catch the few whose
apparent brightnesses change in a systematic, achromatic way due to
microlensing by a nearer star.  The MACHO (MAssive Compact Halo Objects)
collaboration has been remarkably successful  in detecting
approximately 100 microlensing events toward the galactic bulge over a
two year period. The possibility of detecting at least 350 events
per year with modest upgrades in technology together with a larger dedicated
telescope at a good site (C. Stubbs, 1996, private communication) promises
a means to gather planetary statistics at a reasonably rapid rate.  

Although planets that orbit very close to their stars will not be
detectable during a microlensing event, (The star and planet would act
as a single lens.), the likely situation that most planetary systems
will be dispersed more like our own motivates consideration of a
dedicated microlensing search for planets.  How many planets are
likely to be detected in such a search, what is the nature of the
data set, and what planetary statistics can be obtained?  Here we construct
examples of answers to these questions based on the Gould and Loeb
(1992) probability of detecting a planet given that the planet's
central star is acting as a lens during a microlensing event.  

The source stars are treated as point sources in the derivation of this
probability, which means that the application to small mass planets
is not really valid.  However, Bennett and Rhie (1996) have calculated
detection probabilities given that a microlensing event is taking
place for planet/(stellar lens) mass ratios $m/M=10^{-5}$ and
$10^{-4}$ at distances of 4 and 6.4 kpc and for sources at the galactic
center of radii $3R_\odot$ and $13R_\odot$ as a function of the
planet-star separation.  These detection
probabilities averaged over the mass function of the lenses for the
$3R_\odot$ source radius and $m/M=10^{-5}$ for a set of specific
planetary orbital semimajor axes are remarkably close to the
probabilities for the same fixed mass ratio from the point source
calculation (Table \ref{tab:bravep}). This indicates that our
probabilities of detection for the Earth-mass planets for point
sources may be representative to within a factor of 2.  Even without
this near agreement, our purpose here is
to demonstrate the assumption dependence of the expectations from any
microlensing search for planets, to show the effect of the stellar
mass function where most of the lenses will be spectral class M stars,
and to show how well a minimum data set can be interpreted.  The
variations in the number of planetary detections due to a variation in the
assumptions is at least comparable to the errors in the estimates due
to considering only the consequences for point sources. 

We begin in Section 2 with a brief description of the simplest microlensing
theory, and a description of the Gould and Loeb (1992) derivation of
the probability of detecting a planet during a lensing event sufficient to 
allow one to understand the scaling of the published probability to  
arbitrary planet/star mass ratios and arbitrary mean lens
distances. The distribution of events over a lens mass function like
that of the stars in the local neighborhood is developed in Section 3,
and it is used to average the scaled probability of detection of a
planet of given mass, orbit semimajor axis, and average distance of
the lens from the observer. 

A sample observing program like that
advocated in the Roadmap for the Exploration of Neighboring Planetary
Systems (Tytler, 1995) is described in Section 4 after
showing how the planet/star mass ratio and the planet-star separation
normalized by the Einstein ring radius of the lens star is deduced
from the microlensing light curve perturbed by the planet.  Two color,
high time resolution photometry of all ongoing microlensing events
with 1\% accuracy is advocated in this program.  Two colors are
necessary for redundancy, for distinguishing intrinsically varying
stars, and to reduce uncertainties in some cases where the source
star is resolved by the planet. The 
probability of being able to observe the high density star region in
the galactic center (GC) from each of four specific observing sites at any
specific time is used to develop the overall probability of observing
the GC from at least one of the observatories as a function of the time of
year. This is averaged over the time the GC is observable to show that
this particular
distribution of telescopes for a microlensing search could cover the
ongoing events with high time resolution photometry about 54\% to 62\% of
the time.  

A reasonable data base of 3000 events 
assumed collected over an 8 year period is used in Section 5 to determine the 
number of planets that would be detected for three sets of assumptions about 
systems similar to the solar system. For one of these systems a sample 
planetary data set 
consisting of the planet/star mass ratios and planet-star separations 
is constructed from the appropriate probability distributions. An 
interpretation of this data set follows in Section 6 where it is assumed that 
only information from the detailed light curve is available with 
spectral classification of the brighter lenses used to further constrain 
these lens masses.  Although only the planet/star mass ratios follow from the 
light curves, the limited range of lens masses means a reasonable estimate of 
the distribution of planetary masses is obtained.  Observational selection 
effects make the orbital semimajor axes of the planets less defined 
except for the general trend relative to the Einstein ring radius that 
reflects the assumptions about the nature of the systems. 

In Section 7
we show how the finite sizes of the stellar sources affects the
detection of Earth sized planets, where one gains because of the
higher probability of the small planet affecting the light curve and
longer duration of the event, but loses because of the much lower
amplitude of the event.  We also show examples of averaging the
Bennett and Rhie (1996) detection
probabilities (for finite sized sources) over the lens mass function,
compare these probabilities to those obtained for point sources and
point out the necessary steps for obtaining an overall probability of
detecting small mass planets while accounting for the finite source
sizes of the galactic bulge stars.

\section{Microlensing and planet detection}

\begin{figure}
\plotone{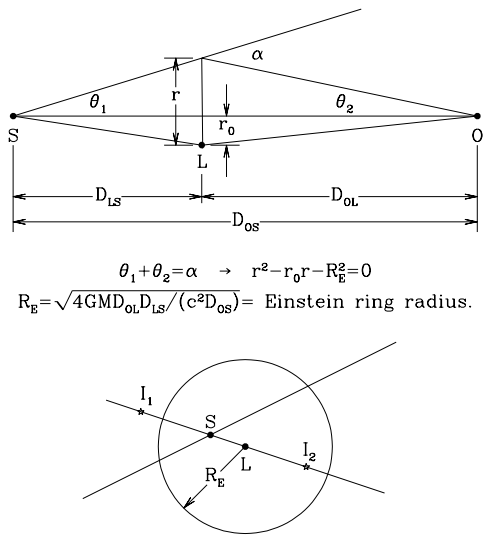}
\caption{Geometry of a microlensing event and definition of
parameters. The circle in the lower diagram is the Einstein ring of
the lensing star and $I_{\s 1}$, and $I_{\s 2}$ are the images of the
source S. O, L and S refer to Observer, Lens and Source respectively. 
\label{fig:geom}}
\end{figure}

\begin{figure}
\plotone{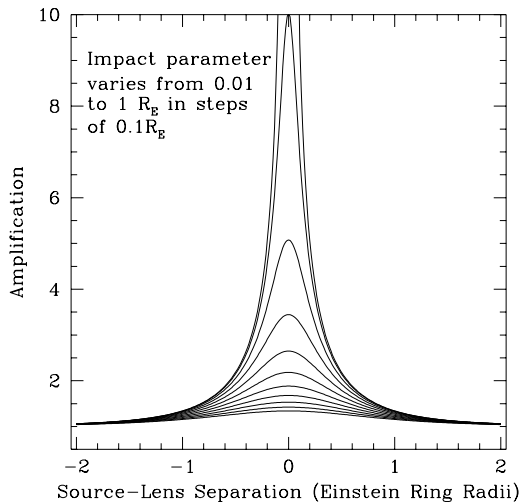}
\caption{Microlensing light curves for several impact parameters of the
relative motion of source and lens stars.  The independence of the
time scale of the event on the impact parameter is
illustrated. \label{fig:ltcvs}} 
\end{figure}

The term microlensing comes from extra-galactic astronomy where the
images of a quasar lensed by a single star in an intervening galaxy
are separated by the order of microarcseconds (Chang and Refsdal, 1979). The
images of a star in the galactic bulge that is lensed by an
intervening star are separated by milliarcseconds, but the term is
still applied. Fig. \ref{fig:geom} shows the geometry of the
gravitational lensing of a point source and the definition of several
parameters.  O, L and S refer to observer, lens and source
respectively. The angle $\alpha$ is the general relativistic (small angle)
deflection of a ray from the source that reaches the observer. The
point source lens equation deduced from the figure is $r^2-r_{\s 0}r
-R_{\s E}^2=0$, where 
\begin{equation}
R_{\s E}=\sqrt{\frac{4GMD_{\s OS}z(1-z)}{c^2}},\label{eq:eringr}
\end{equation}
is the Einstein ring radius. The Einstein ring
is the symmetric image of the source when it is directly behind the
lens ($r_{\s 0}=0$). Here $G$ is the gravitational constant, $M$ is
the mass of the lens, $c$ is the velocity of light and $z=D_{\s
OL}/D_{\s OS}$. 
The circle centered on L in Fig. \ref{fig:geom}
is the Einstein ring on the lens plane. Solution of the lens equation
$r_{\s 1,2}=(r_{\s 0}\pm\sqrt{r_{\s 0}^2+4R_{\s E}^2})/2$ yields the  
positions of the two
images $I_{\s 1}$ and $I_{\s 2}$ outside and inside the Einstein ring
respectively.  The amplification ({\it e.g.}, Gould and Loeb, 1992),
\begin{equation}
A=\frac{u^2+2}{u\sqrt{u^2+4}},\label{eq:amp}
\end{equation} 
is the factor by which the flux density from the source star is
increased. It is found for each image as the transformation of areas
from source to image coordinates with the results for both images
being added together.  Here $u=r_{\s 0}/R_{\s E}$. The light curve of the event
reaches a maximum when $u$ has its minimum value of $u_b$. The time
scale of the event is the time for the projected position of the
source to move an Einstein ring diameter, $\tau=2R_{\s
E}/v\propto\sqrt{M}$ where $v$ is the relative speed of the source
projected onto the lens plane.  The singularity in $A$ at $u=0$ is
called the caustic point.  This singularity is removed when the source
is not a point source. Fig. \ref{fig:ltcvs} shows the light curves
for a variety of impact parameters and illustrates the independence of
the perceived time scale on the latter.

If the lens is a relatively close binary system, the lens equation
relating coordinates of the images in the lens plane to the coordinates
of the source in the source plane is now a vector equation (Schneider
and Weiss, 1986). The quadratic lens equation for a single lens becomes
two 5th order equations or a single fifth order equation in the
complex plane (Witt, 1990). There are now either three or five images
and the amplification is calculated for each image as before as the
transformation of areas with the results summed for the total. The
caustic point for the single lens is now one or two closed caustic
curves in the source plane that transform into critical curves in the
lens plane, where $A\rightarrow \infty$ when the point source is on a
caustic.  The singularity again vanishes for finite sized
sources. There are three images when the source is outside a closed
caustic curve and five when it is inside.  A sharp peak in the light
curve occurs when the source crosses a caustic with increased
amplification while the source is inside.  The simple, bell shaped
light curve for a single lens can become quite complicated for a
binary lens but also quite informative ({\it e.g.} Alard, {\it et al.}
1995).  

\begin{figure}[t]
\plotone{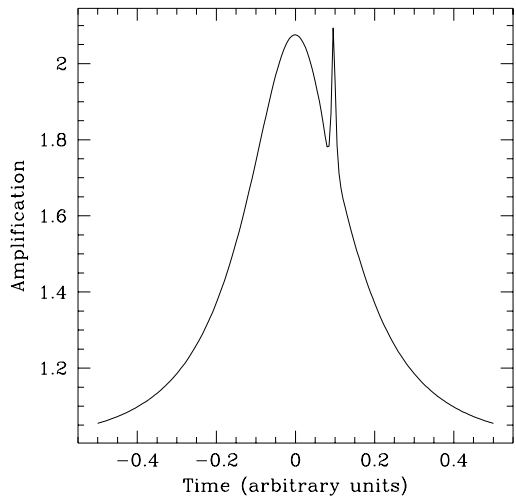}
\caption{Example of a microlensing light curve perturbed by a planet. The
planet/star mass ratio $m/M=0.001$ and the planet is located $1.3R_{\s
E}$ from the lens. (After Gould and Loeb, 1992)\label{fig:planltcv}}
\end{figure}

When the second member of the binary system is a planet, the small
mass of the planet compared to the star leads to a 
significant deviation of the light curve from that of the star without
the planet only if the one of the two unperturbed images of the
source comes close to the planet's projected position in the lens plane
(Gould and Loeb, 1992).  Fig. \ref{fig:planltcv} shows an example
of such a perturbed light curve, where the planetary signature is 
larger than the simple sum of the gravitational lensing of the star
plus the planet because the planet is focusing light already focused
by the central star.  For an event involving a lens with a 
planetary companion, the planet can affect the light curve of a point
source by at most
a very short fraction of the total time of the event as shown in
Fig. \ref{fig:planltcv}, and for most events, the planet will not
reveal its
presence. However, for a point source near the galactic center being
lensed by a solar mass star at 4 kpc from the Sun, the probability that
a Jupiter mass planet 5.2 AU from the star will cause at least a 5\%
deviation from the unperturbed light curve some time during the event
is about 0.17 (Gould and Loeb, 1992).  
(An event is said to occur when the lens and source
come within an Einstein ring radius of the lens of each other.)  This high
probability results from the semimajor axis being near the
Einstein ring radius of the star where the probability is maximized,  
but there is a considerable range of
semimajor axes ($2.4\,AU\leq a\leq 7.3\,{\rm AU}$ for this case) where the probability
exceeds 0.1 (Gould and Loeb, 1992). David Tytler (private
communication, 1995) has coined the term ``lensing zone'' for this
range of semimajor axes where the probability of detection is high. 

\begin{figure}[t]
\plotone{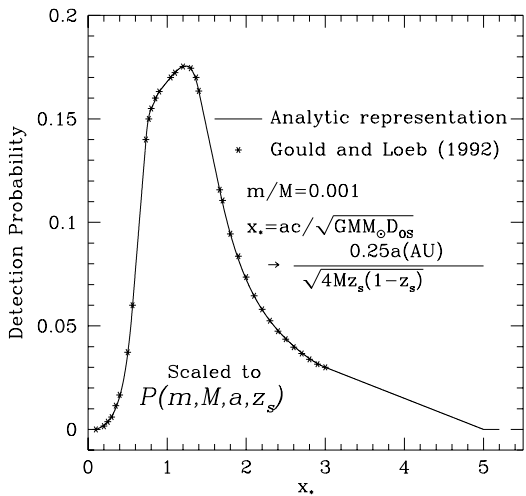}
\caption{Analytic representation of the probability of detecting a
planet with $m/M=0.001$ given that an event is taking place, where a
point source is assumed and where a 5\% perturbation of the light
curve anytime during the event is assumed detectable. The curve has
been linearly extended beyond the Gould and Loeb data. The second form
of $x_*$ is to account for the scaling of the position of the lensing
zone for various average fractional distances of the lens $z_s$ and
lens masses $M$. ($M$ is normalized by $M_\odot$.)
\label{fig:glprob}}
\end{figure}

An analytic representation of the Gould and Loeb (1992, Fig. 8) 
detection probability for a planet/star mass ratio $m/M= 0.001$ and
for a 5\% detection threshold (Appendix A) 
is shown in Fig. \ref{fig:glprob},
where $x_*=ac/\sqrt{GMD_{\s OS}}$ is the scaled semimajor axis, and
where the curve has been linearly extended for $(3\leq x_*\leq 5)$.
We shall use this probability of detection in generating a sample data
set for a microlensing planet detection program, but several
qualifications and extensions are necessary.  First the curve was
derived under the assumption that the sources are point sources. The
probabilities of detection are thus representative only as long as the angular
size of the Einstein ring of the planet is large compared with the
size of the source.  This follows because a point source must
be inside the Einstein ring of a lens for significant amplification
and therefore only a fraction of a finite source can be amplified by a
small lens with a consequently smaller perturbation of the light curve.  At
the same time, there is a larger probability that some part of the now
finite sized image of the source will encounter the planet in the lens
plane.  Recent calculations by Bennett and Rhie (1996) have
demonstrated that this enhanced probability of planet encounter may more
than compensate for the lower amplitude of the perturbation in the
sense that their maximum probability of detection is about twice the
maximum probability of detection for a point source with a planet/star
mass ratio of $10^{-5}$.  Averaging the Bennett and Rhie detection
probabilities over the lens mass function in Section 7 yields
probabilities of detection close to those obtained with a point
source, so small mass planet detections determined with the point
source assumption should be within a factor of 2 of those accounting
for the finite source size.  

Next, the Gould and Loeb probability
was derived by averaging over a distribution of lenses
along a particular line of sight that was displaced about $4^\circ$
from the direction to the galactic center yielding a uniform
distribution of lenses. Other lines of sight for
the assumed spatial distribution of stars in the galaxy would yield
different probabilities, but not significantly so.  The fact that the
number of detected events toward the galactic center was about three
times that expected from the distribution of stars assumed in the
Gould and Loeb analysis (Alcock, {\it et al.} 1995) will affect the  
probabilities of detection by moving the position of the average lens
closer to the center of the galaxy thereby reducing the size
of the Einstein ring radii (Eq. (\ref{eq:eringr})).  This follows
since the excess lenses are thought to be located in a bar-like
distortion of the galactic bulge distribution of stars oriented
towards  the Sun (Zhao {\it et al.} 1995). An approximate correction
for this effect is made below.

The correction in the probability for moving the
average lens closer to the source as well as scaling the probability for
different values of $m/M$ will be better understood if we outline how the
probability of detection is determined.  If $m/M\rightarrow 0$, the
binary lens solutions go to the single lens case with the two unperturbed
images described above. For small $m/M$ only one of the images will be
significantly affected when the planet approaches a separation from
the unperturbed image position comparable to planet's own Einstein ring
radius, and the new images will be close to the unperturbed position
of the image. The amplification is calculated as before as
the sum of the amplifications of each image. As the planet must be
close to the unperturbed image position for significant perturbation
of the light curve, contours of planetary
positions for a given percentage perturbation of the light
curve will be grouped close to the unperturbed image position with the
percentage perturbation decreasing for contours further away from
that position.  

In Fig. \ref{fig:geom} we can imagine a contour
of the planetary position in the lens plane corresponding to a 5\%
perturbation surrounding each of the two nonperturbed image positions.
Examples of these contours are shown in Fig. 2 of Gould and Loeb
(1992). If the planet is inside one of these contours drawn in the
lens plane, the perturbation of the light curve will exceed 5\%. 
As the source and lens pass, the unperturbed images describe arcs in
the lens plane (see Fig. \ref{fig:geom}), and the loci of the
contours as the images change their positions define two areas in the
lens plane. If the projected position of the planet is inside either of
these  areas,
there will be a perturbation of the light curve exceeding 5\%
sometime during the event and the planet will be detectable.  The
probability of the planet detection sometime during the event is just
the probability
that the planet is within either of the two areas.  Finally the
probability is determined for and averaged over all impact parameters
less than the Einstein ring radius.

The first step in determining this probability is to assume that a
planet with semimajor axis $a$ has a uniform probability distribution
over a sphere of radius $a$ to account for random orbit inclination
and phase. Then the probability that the planet has a projected
separation between $r$ and $r+dr$ is 
\begin{equation}
p(r)dr=\frac{rdr}{a^2\sqrt{1-r^2/a^2}},\label{eq:pdense}
\end{equation}
and the probability that the planet is between $r$ and $a$ is
$F(r)=\sqrt{1-r^2/a^2}$.  If we normalize the
distances by the Einstein ring radius $R_{\s E}$ from
Eq. (\ref{eq:eringr}),
\begin{equation}
F(x_p,z)=\left[1-4z(1-z)\frac{x_p^2}{x_*^2}\right]^{1/2}\label{eq:Fxpz}
\end{equation}
represents the probability that the planet will be located between
$x_p=r/R_{\s E}$ and $a/R_{\s E}$ from a lens located at
fractional distance $z$ to the source with $x_*=ac/\sqrt{GMD_{\s
OS}}$ as before. Gould and Loeb (1992) average Eq. (\ref{eq:Fxpz}) over the
line of sight weighted by the probability that an event will occur to
obtain
\begin{equation}
F(x_p)=\frac{\int_0^{1/2-z_{\s 0}}\cdots dz+\int_{1/2+z_{\s 0}}^1R_{\s
E}(z)\nu(z)F(x_p,z)dz}{\int_0^1R_{\s E}(z)\nu(z)dz}\label{eq:Fxp},
\end{equation}
where $\nu(z)$ is the spatial density distribution and
where $z_{\s 0}=0$ if $x_p<x_*$ and $z_{\s 0}=0.5\sqrt{1-x_*^2/x_p^2}$
if $x_p>x_*$. Although this is the form used by Gould and Loeb in their
calculations, it differs from their Eq. (3.8) in that the 
integral is separated into two segments.
The split limits are necessary since the integrand becomes
undefined when $x_p>x_*$. At $z=1/2$, $x_p>x_*$ implies
$r>a$ which is clearly impossible. As $z$ moves away from 1/2,
$x_p=r/R{\s E}$ gets larger as $R_{\s E}$ gets smaller.  So to have a
value of $r$ still inside the maximum planetary separation from the
star for larger $x_p$, the integration can extend only over those
values of $z$ where large ${x_p}$ still corresponds to $r<a$.

The particular form of $\nu(z)$ chosen by Gould and Loeb places the
average lens near 4 kpc from the Sun. If we let
$\nu(z)=C\delta(z-z_s)$ where $C$ is some constant, all of the lenses
are at $z=z_s$ and
$F(x_p)=F(x_p,z_s)=\sqrt{1-4z_s(1-z_s)(x_p/x_*)^2}$. The probability
distribution 
\begin{displaymath}
-\frac{dF(x_p)}{dx_p}=\frac{4x_pz_s(1-z_s)}{x_*^2\sqrt{1-4z_s(1-
z_s)(x_p/x_*)^2}} 
\end{displaymath}
is strongly peaked at a removable singularity at
$x_*=x_p\sqrt{4z_s(1-z_s)}$. This is where the probability of finding
the planet per unit area in the lens plane would be maximal.  Now
the contours surrounding the unperturbed image positions within which
the perturbation to the light curve exceeds 5\% are maximal in size
when the image comes close the the Einstein ring and the locus of
contour areas during the event is likewise maximized. As the planet must
be near the image to be inside the contours, $x_p\sim 1$ ({\it
i.e.} planet near the Einstein ring). The combination of the largest
areas in the lens plane for which the perturbation of the light curve
would exceed 5\% sometime during the event, if the planet were inside
the area, with the maximum probability per unit area corresponds to the
highest probabilities for finding the planet. So for $x_p$ near 1,
$x_*\sim \sqrt{4z_s(1-z_s)}$ corresponds to the peak in the probability
curve.  For $z_s=1/2$ the peak should be near $x_*=1$, and we see from
Fig. \ref{fig:glprob} that placing all the lenses at $z=1/2$ gives a
peak in the probability of detection near the peak that was obtained 
for the average
lens being at $z=1/2$.  So for the average lens being at a value of
$z$ other than 1/2, we shift the lensing zone by $\sqrt{4z(1-z)}$. This
is equivalent to writing $x_*=0.25a/\sqrt{4Mz_s(1-z_s)}$ in reading the
probabilities from Fig. \ref{fig:glprob}, where $a$ is in AU, $M$ is
in units of $M_\odot$, $z_s$ here is approximately the value of $z$ for
the average lens, and other constants have been evaluated. 

Next we must scale the probability of detection to planet/star mass
ratios $m/M$ other than the value of 0.001 appropriate to Fig.
\ref{fig:glprob}. We note that the size of the 5\% contour around the
unperturbed image position in the lens plane scales as 
$\sqrt{m/M}$ (Gould and Loeb, 1992). As the arc described by
unperturbed images is long compared to any dimension across the
contours and this arc is independent of $m/M$, only the dimension
of the area of the locus of contours perpendicular to the image arc is
affected by a change in $m/M$, so
the area of the locus of 5\% contours for the complete event scales
as $\sqrt{m/M}$. Since the probability per unit area of the planet
falling into the area of the locus of contours does not vary much over
the area, the probability is changed approximately as the area is
changed. Hence, for a given value of $x_*$, we need but multiply
the probability by $\sqrt{m/0.001M}$ to obtain the probability of
detection for an arbitrary planet/star mass ratio. We will thus
represent the Gould and Loeb probability curve in
Fig. \ref{fig:glprob} so scaled and shifted as $P(a,m,M,z_s)$.  This
is the probability, during a lensing event, of detecting a planet of
mass $m$, with semimajor axis $a$, orbiting a lens star of mass $M$
located at a mean fractional distance $z_s$.  Below
we will average this probability over the mass function for particular
sets of values of $a,m,z_s$, but first the distribution of events
over the mass function must be developed. 

\section{Distribution of the events over the lens mass function}

We shall assume that the mass function for local stars is
representative of the stars between us and the center of the
galaxy. This assumption will probably be most in error for masses
larger than about $1.5M_{\sun}$, since more massive lenses within the
galactic bulge will have evolved off the main sequence and thereby
will have lost mass on their way to becoming white dwarfs or neutron stars.
The mass function used here is an analytic approximation derived 
from the multiplicity corrected present day mass function for main
sequence stars given in
Table 1 and Fig. 5 of Basu and Rana (1992). We consider only the
range of masses $0.08\leq M\leq 2.0$ because of the very small
fraction of stars of higher mass and because of the completed evolution of the
older possible lens stars in the near side of the galactic bulge. M is
expressed in units of the solar mass $M_{\sun}$.
This mass function of main sequence stars neglects the contribution of
white dwarfs and neutron stars to the lens population, but these
latter stars are expected to be a small fraction of those stars
remaining on the lower main sequence.  The mass function is
represented by (based on Basu and Rena, 1992)
\begin{eqnarray}
\phi(M)&=&48.39M^{-1}\qquad 0.08\leq M\leq 0.5327,\nonumber\\
       &=&13.73M^{-3}\qquad 0.5327\leq M\leq 1.205,\nonumber\\
       &=&28.04M^{-6.83}\qquad 1.205\leq M\leq 2.0, \label{eq:massfn}
\end{eqnarray}
where $\phi(M)dM$ is the number of stars/$\rm pc^2$ in the solar
neighborhood in the mass range $dM$ about $M$ integrated perpendicular
to the galactic plane.

The fraction of stars
in the range $dM$ about $M$ is found by dividing $\phi(M)dM$ by the
integral of $\phi(M)dM$ over $0.08\leq M\leq 2.0$, but this is not the
fraction of microlensing events to be expected for lenses in mass
increment $dM$.  This follows from our definition of an ``event'' as
the lens and source coming within an angular radius of each other
equal to the Einstein ring angular radius $R_{\s E}/D_{\s OL}$ of the
lens.  The dependence of $R_{\s E}$ on $M$ (Eq. (\ref{eq:eringr})) means
that a more massive lens has a greater probability of having a source
pass within its Einstein ring radius than a less massive lens.  

To evaluate the distribution of events over the mass function, we
assume first for simplicity that $D_{\s OS}=8\,{\rm kpc}$ is the same
for all the source stars in the galactic bulge, so that $z=D_{\s
OL}/D_{\s OS}$ is the fractional distance of the lens to the center of
the galaxy.  If $v_{\s\perp}(z)$ is the magnitude of the component of
velocity of a single lens  perpendicular to the line of sight, and $N$
is the number of sources per unit solid angle toward the galactic
bulge, $Nv_{\s\perp}(z)2R_{\s E}(M,z)/(z^2D_{\s OS}^2)$ is the number
of encounters (events) per unit time experienced by the single lens of
mass $M$ at fractional distance $z$. Next, let $\nu(r)$ be the number
density of stars at distance $r$, such that $\nu(r)dr$ is the areal
density of stars for the range between $r$ and $r+dr$. Let $f(M)dM$ be
the fraction of stellar masses between masses $M$ and $M+dM$, which is
constructed from $\phi(M)$ (Eq. (\ref{eq:massfn})) as described
above. Setting $r=D_{\s OS}z$, we have $f(M)dM\nu(z)D_{\s
OS}^3z^2dz$ is the number of stars between $M$ and $M+dM$ per unit
solid angle that are in the slab between $z$ and $z+dz$.  Multiplying
this by the events per unit time due to a single lens of mass $M$ at
$z$ and inserting the explicit $M$ and $z$ dependence into $R_{\s E}$ yields
\begin{equation}
2\sqrt{\frac{4GM}{c^2}}D_{\s OS}Nf(M)dM\int_0^1\nu(z)\langle
v_{\s\perp}(z)\rangle\sqrt{z(1-z)}dz  \label{eq:evtsdens}
\end{equation}
as the number of events per unit solid angle per unit time due to
lenses between $M$ and $M+dM$ over the entire line of sight. Here
$\langle v_{\s\perp}\rangle$ is an average transverse
speed. Then from Eq.(\ref{eq:evtsdens}),
\begin{equation}
{\cal F}(M)dM=\frac{\sqrt{M}f(M)dM}{\int_{0.08}^{2.0}\sqrt{M}f(M)dM}
\label{eq:evtsdist} 
\end{equation}
represents the fraction of events due to lenses with masses between $M$
and $M+dM$, where common factors from Eq. (\ref{eq:evtsdist}) 
have been cancelled in numerator and
denominator.  The probability of detection of a planet of mass $m$,
semimajor axis $a$, located at average fractional distance $z_s$
weighted by the distribution of events over the mass function along
the line of sight is 
\begin{equation}
P(m,a,z_s)=\int_{0.08}^2{\cal F}(M)P(m,M,a,z_s)dM. \label{eq:pave}
\end{equation}
The distribution of this probability over the $m,a$ plane is shown in
Fig. \ref{fig:pmaplane} for the case where the planetary mass $m$ and
semimajor axis $a$ are invariant over the lens mass function.  It
is used  in section 5 to deduce sample results 
from a dedicated microlensing search for planets, but first in the following 
section we define a representative observing program that accounts for 
observing constraints.
 
\section{Sample Observing Program}

Recall that the planet-image separation must be on the order of the
Einstein ring radius of the planet or less to cause a significant perturbation
of the single lens light curve.  The time scale for the whole event is
the time required by the source to traverse an Einstein ring diameter of
the star (see Fig. \ref{fig:ltcvs}). The time scale of the planetary
perturbation is the time
required for the image to traverse the Einstein ring diameter of the
planet, which is comparable to the time required for the source to
traverse the Einstein ring diameter of the planet.  The relative
transverse speed of the
source is the same for both cases, so the ratio of the time scale of
the planet perturbation to the time scale of the event is just
$\sqrt{m/M}$ (see Eq. {\ref{eq:eringr}).  If an event perturbed by a
planet with mass ratio 0.001
is observed to last 45 days, the planet perturbation would last only
about 1.4 days.  For a point source, a perturbation by an Earth mass planet
about a solar type star ($m/M=3\times 10^{-6}$) would last only about
1.8 hours for the same 45 day main event.  However, the effect of the
finite size of the source on the detection of small mass planets is
profound, where a perturbation of the light curve is both reduced in
amplitude and extended in duration beyond the 2 hours expected for a
point source.  For this reason, the observing
strategy must be fundamentally different for detecting the small mass
planets. We will discuss this in more detail in Section 7.  
If a 5\% perturbation of a
light curve is considered detectable, all events would have
to be photometrically monitored with about 1\% accuracy at time
intervals no longer than about one hour to detect a 
perturbation by Earth-mass planets and if a perturbation is detected, 
at intervals of a few minutes thereafter on that particular light
curve for the duration of the perturbation.  In reality, the time
scale for the planetary perturbation for a given mass ratio will vary
somewhat depending on the angle that the source crosses the
given magnification contours in the source plane. But that angle is
known from the position of the planetary perturbation on the
unperturbed light curve. This defines the intersection of the source
trajectory with the line connecting the lens and planet. From this
angle and an accurate, high time resolution light curve obtained
during the perturbation, one can model the
light curve and refine the determination of the mass ratio.

The ratio of the planet-star separation and the Einstein ring radius
$r_p/R_{\s E}$  also follows from the position of the planetary 
perturbation on the unperturbed light curve. If the perturbation is removed, 
the unperturbed amplification yields $u=r/R_{\s E}$ from Eq. (\ref{eq:amp}).
For each value of $u$, the positions of the two unperturbed images are known 
and the planet must be near one or the other image to have perturbed the 
light curve.  The selection of the correct image between the two can
follow from the modeling of the light curve. If
the caustic in the source plane is intercepted by the point source,
the planetary position is known to within the Einstein ring radius of
the planet.

David Tytler (1995) has suggested a
telescope distribution and observing program to accomplish the necessary
observations.  One of four two meter telescopes at established
sites distributed as widely as possible in longitude in the southern
hemisphere would monitor several tens of millions of stars in the
galactic center (GC) in two colors at least once a night whenever the bulge
is observable.
This ``survey telescope'' would discover as many microlensing events
as possible using a technique that is currently being used
successfully by the MACHO collaboration (Alcock {\it et al.} 1995, 1996).
The three ``followup telescopes'' would photometrically monitor each
of the ongoing microlensing events in two colors at the high accuracy 
and high time resolution described above.  We have arbitrarily added a
fourth followup telescope in Hawaii.

\begin{table*}[t]
\centering
\begin{tabular}{cccc}
\hline
Observatory&Latitude (deg.)&E. Longitude (deg.)\\
\hline
LaSilla&-29.25&289.27\\
Siding Springs&-31.27&149.06\\
South Africa&-32.38&20.80\\
Hawaii&+19.83&204.52\\
\end{tabular}
\begin{center}\caption{Location of selected observatory sites.
\label{tab:obsloc}}\end{center}
\end{table*}
The locations of three developed observatory sites in the southern
hemisphere separated maximally in longitude plus Hawaii in the northern
hemisphere are shown in Table \ref{tab:obsloc}. 
Weather, equipment problems and maintenance prevent 100\% coverage of
the events with high precision, high time resolution photometry. The
fraction of usable nights (spectroscopic) at each observatory site is
shown in Fig. \ref{fig:wea} where the curves are constructed from
\begin{figure}[t]
\plotone{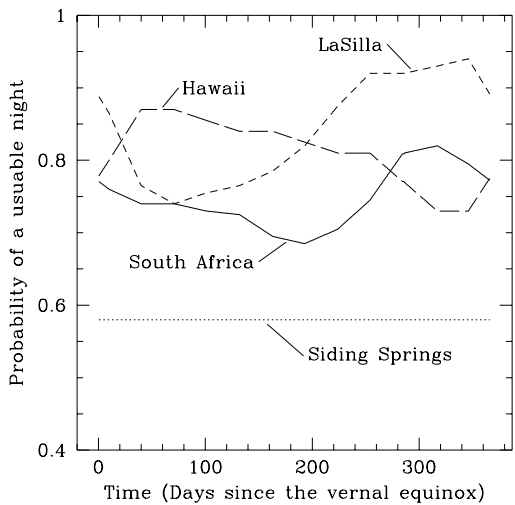}
\caption{Probability of a usable night as a function of the time of
year for four observatories that are suitable locations for dedicated
telescopes in a microlensing search for planets. \label{fig:wea}}
\end{figure}
monthly averages for LaSilla (http://lwh1.ls.eso.org) and South Africa
(J. Caldwell, Private communication, 1996), quarterly averages from
Hawaii  (J. Glaspey, Private communication, 1996), and yearly averages
from Siding Springs where summer and winter are comparable (J. Mould,
Private communication, 1996).  The GC can only be observed
from a given site at some time $t$ when it has a zenith angle less than
$70^\circ$ while the solar zenith angle is greater than
$105^\circ$. If these criteria are satisfied, the probability that the
GC is observable at the particular time is taken from
Fig. \ref{fig:wea}.  Otherwise the probability of observing the
galactic center from that site at that time is zero.  

The probability  that the GC
is observable from at least one of the observatories at any
instant is then just the union ({\it e.g.} Feller, 1957, p. 88)
of the probabilities for each site at that
instant.  This union of probabilities
differs from that for one or another of the sites only when the GC
is simultaneously observable from more than one site. The sum
$\sum_{i=1}^nP(t_i)\Delta t_i$ gives the fraction of a day that the
GC can be observed, where $P(t_i)$ is the probability of observation at
time $t_i$ and $\Delta t_i=1/n$ is an increment of a day with $n$ being
large but otherwise arbitrary. The
probability of being able to observe the galactic center on a given day
so calculated 
and thus to make photometric measurements of lensed stars is shown in 
Fig. \ref{fig:probobs} as a function of time. The maximum observability
\begin{figure}[t]
\plotone{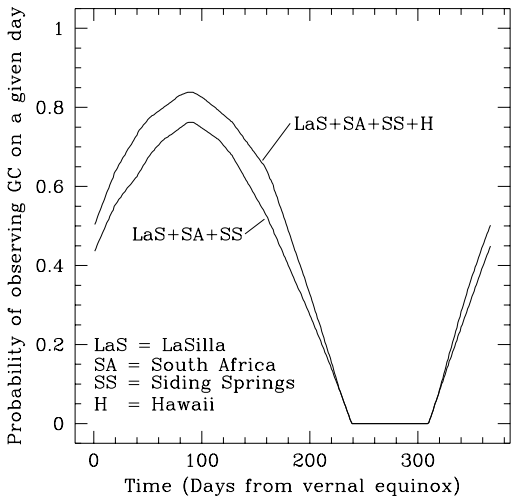}
\caption{Probability of observing the galactic center (GC) on a given day.
A probability of 1 would indicate that the GC was observable at all
times during the day from one or more observatories.  The GC is
considered observable if the zenith angle of the GC is less than
$70^\circ$, the zenith angle of the Sun is more than $105^\circ$ and
the night is usable. \label{fig:probobs}}
\end{figure}
occurs in the southern winter and drops to zero when the Sun is near the
position of the galactic center during southern summer.  There would be
a period of time spanning day 90 in Fig. \ref{fig:probobs} where the
probability of observing the GC would be unity if the weather were
always perfect and equipment always worked, since the GC would 
always be observable from at least one of the sites throughout the
day. The addition of Hawaii to the three southern sites gives a significant
increase in the probability of observation because of its low northern
latitude and because of its ideal location in longitude between Siding
Springs and LaSilla. 

As the Sun approaches the position of the galactic bulge in the sky,
the duration of dark sky while the GC is up shrinks to the point where
all of the ongoing events cannot be covered with high precision
photometry in the time available.  In fact there is a point beyond which
the survey telescope cannot cover all of the fields in its program,
and observations would probably cease for a time longer than the 71 days
when the GC is completely blocked by the Sun in
Fig. \ref{fig:probobs}.  The MACHO group requires about an hour to
cover 24 fields toward the galactic bulge with an additional hour of
overhead (Alcock, {\it et al.} 1966; C. Alcock, private communication,
1996). If we assume that two hours is the minumum amount of dark time
for observing the GC, the 71 day exclusion region in
Fig. \ref{fig:probobs} is expanded to about 120 days, leaving 245 days
as the maximum span of time per year for observing the GC in a microlensing
survey and for followup photometry. If we center the 245 day span over
the peak probability in Fig. \ref{fig:probobs} and average the
probability over this span of time, we obtain the average fraction of the time
when the galactic center can be monitored with high time resolution
photometry during observed microlensing events. This fraction of
0.54 for just the three southern observatories is increased to 0.62 with
the addition of a fourth telescope in Hawaii.   

The approximately 58\% average coverage for the assumed configuration of
telescopes means that a little less than half of the planetary
perturbations of 
microlensing light curves {\it of very short duration} will be
missed. However, if the duration of the perturbation to the light
curve exceeds a day, as it would in most cases for a Jupiter mass
planet, the curves in Fig. \ref{fig:probobs} represent the fraction
of the perturbation that could be monitored on the average where the
probability of detecting the perturbation is essentially unity.  This
latter assertion follows since the probability of {\it all} of the
sites having unusable nights on a given day is  
\begin{displaymath}
P_{\rm all\,cloudy}=\prod_i(1-P_i)\approx 3\times 10^{-3}\,{\rm
to}\,3\times 10^{-4},
\end{displaymath}
where $P_i$ is the probability of a usable night at the $i$th
observatory taken from Fig. \ref{fig:wea}, and where the range of
values results from the variation in the weather conditions during the
year.  The fraction of the longer perturbations of the light curve that must
be followed for meaningful interpretation is not large if the groups
of high precision points are well distributed as they normally would
be as the event is passed from one observatory to the next.  

A few more of the small planets will be missed because of abandoning complete
coverage to take measurements of ongoing perturbations every few
minutes, but these will be neglected in the following.  The
probability of observing the GC in Fig. \ref{fig:probobs} would
approach unity for a time spanning the southern winter if an arbitrary
number of followup telescopes could be distributed more or less
uniformly in longitude near $-30^\circ$ latitude ({\it e.g.}, see
Gould and Loeb, 1992).  The impracticality of developing many new
sites and necessary ocean gaps preclude this ideal. 

Alcock {\it et. al.} (1996) found 43 likely microlensing events in the first
year galactic bulge data where 12.6 million stars were monitored for
190 days. For the same detection efficiency, following a little more than 100
million stars would allow the detection of 350 events per year.  However,
from the Alcock {\it et al.} data, the optical depth for microlensing
events for the
distance from the Sun to the center of the galaxy is
$3.9^{+1.8}_{-1.2}\times 10^{-6}$, which is based on the 13 lensed 
giants that are concentrated near the galactic center.  An overall
optical depth, based on the entire data set, was estimated to be
$2.43^{+0.54}_{-0.45}\times 10^{-6}$, where the latter number is
thought to be smaller because some of the main sequence sources were
closer to us than the center of the galaxy.  Both optical depths account
for empirical determinations of the detection efficiency, which never
exceeded 0.4 and decreased toward zero as event time scales approached a
few days.  These optical depths imply that far fewer than 100
million stars would have to be monitored to detect 350 
events per year if the detection efficiency were increased by using a
larger telescope at a better site along with only modest improvements
in technology and observational procedures (C. Stubbs, private
communication, 1996). If each event lasts an average of about 40 days
(Alcock, {\it et al}, 1996), 350 events in a 190 day observing
period would mean that almost 75 stars would be lensed at any one time.  
With continued improvements in technology and
observing procedures, the observation of about 3000 microlensing events 
over about an 8 year period is not an unreasonable goal. These events
would be covered with precise, 
high time resolution photometry about 58\% of the time on average.  
We will adopt
these numbers in the examples to follow, where the number of detected
{\it small} planets scales directly both with the total number of
events and with the percentage of photometric coverage and the number
of large planets with just the total number of events.

\section{Sample expectations from a microlensing search}

What is expected from any microlensing search for planets depends
entirely on what is assumed about the nature of likely planetary
systems. It is guaranteed that the distribution of planet properties
we find is going to be unexpected, but to illustrate some examples, and to 
show the assumption dependence of the expectations, we shall base our model 
systems on our own solar system.  
In about half of the 3000 assumed events, the lens will 
be a member of a binary system although this binary nature will usually not 
be revealed.  (Close binaries will act as a single lens and binaries
separated by more than twice the Einstein ring radius of the more
massive member will [usually] act as two separate lenses. Attempts to
obtain a secure
distribution of binary separations are frustrated by severe
observational bias (D. Popper, private communication, 1996).)
To be conservative, we assume that none of the binary systems 
have any planets, but to be equally optimistic we assume that all of the 
remaining 1500 events involve lenses with planetary systems similar to
our own, but with 
only four or five planets. 

All 1500 remaining lenses have a Venus 
($m/M_\odot=2.5\times 10^{-6}$) at 0.7 AU, and an Earth ($m/M_\odot=3\times 
10^{-6}$) at 1 AU, but only half or 750 lenses have a Jupiter ($m/M_\odot= 1 
\times 10^{-3})$ at 5 AU and a Saturn ($m/M_\odot=3\times 10^{-4}$) at
10 AU.  The other 750 lenses have Uranuses ($m/M_\odot = 5\times 10^{-5}$) 
at the distances 
of the Jupiters and Saturns and, for these latter systems, an extra Earth at 
2.5 AU. It is
difficult to make a Jupiter in the current planet formation paradigm within 
the observationally estimated lifetimes of preplanetary disks, so we remove 
Jupiter from half of the systems.  The extra Earth at 2.5 AU is 
motivated by Wetherill's (1994) calculations showing the persistence of Earth 
mass planets in the asteroid belt if Jupiter is absent.  Given these 
systems around 1500 of the lenses, how many of the planets would be detected 
in the dedicated microlensing search described above? What is the nature  
of the minimum data set that would be obtained, and how well could it be 
interpreted?  Before attempting to answer these questions, we must make 
assumptions concerning the variation of the system properties with the mass 
of the central star.    

We shall determine the expected number of detected planets for three distinct 
sets of assumptions: 1. The planet-star mass ratios and separations remain  
invariant for all stellar masses $M$. This means that smaller stars have 
smaller planets, but that the Earths, for example, will always be at 1 (or 
2.5) AU regardless of the value of $M$. There is some weak
justification for assuming that the planet-star 
separations are the same regardless of the mass of the primary, at least for 
Jupiters, since the ice condensation point in a preplanetary nebula where 
Jupiters are thought to form may always be near 5 AU (Boss, 1996).  
2. The planet-star mass ratios remain 
invariant, but the separations are always the same fraction of the
Einstein ring radius at $z=0.5$.  Here smaller stars have
smaller planets, but the planets move 
closer to their central stars as $M$ decreases.  3. The masses and 
separations remain invariant instead of the mass ratios and separations.  
In this last case, the Jupiters for example will always have $m=0.001M_\odot$ 
for all values of $M$ and will always be 5 AU from their central
stars.

\begin{table*}[t]
\centering
\begin{tabular}{|c|c|c|c|c|c|c|}\hline
&\multicolumn{2}{|c|}{$m_i/M,\,a_i\sim$const.}&\multicolumn{2}{|c|}{$m_i/M,\,a_i/R_{\s E}\sim$const.}&\multicolumn{2}{|c|}{$m_i,\,a_i\sim$const.}\\\hline
&$z_s=0.5$&$z_s=0.8$&$z_s=0.5$&$z_s=0.8$&$z_s=0.5$&$z_s=0.8$\\
\hline
planet&\multicolumn{6}{|c|}{Probability of detection}\\
\hline
Venus&$6.64\times 10^{-4}$&$1.45\times 10^{-3}$&$6.84\times
10^{-5}$&$1.30\times 10^{-4}$&$1.75\times 10^{-3}$&$3.80\times 10^{-3}$\\
Earth(1)&$2.31\times 10^{-3}$&$3.75\times 10^{-3}$&$2.08\times
10^{-4}$&$3.95\times 10^{-4}$&$5.92\times 10^{-3}$&$8.98\times 10^{-3}$\\
Earth(2.5)&$7.65\times 10^{-3}$&$7.30\times 10^{-3}$&$4.86\times
10^{-3}$&$8.37\times 10^{-3}$&$2.36\times 10^{-2}$&$1.19\times 10^{-2}$\\
Uranus(5)&$1.78\times 10^{-2}$&$1.14\times 10^{-2}$&$3.92\times
10^{-2}$&$3.00\times 10^{-2}$&$2.54\times 10^{-2}$&$1.55\times 10^{-2}$\\
Uranus(10)&$3.49\times 10^{-3}$&$1.68\times 10^{-3}$&$9.80\times
10^{-3}$&$6.29\times 10^{-3}$&$4.30\times 10^{-3}$&$1.89\times 10^{-3}$\\
Jupiter&$7.97\times 10^{-2}$&$5.10\times 10^{-2}$&$1.75\times
10^{-1}$&$1.34\times 10^{-1}$&$1.14\times 10^{-1}$&$6.95\times 10^{-2}$\\
Saturn&$8.65\times 10^{-3}$&$4.11\times 10^{-3}$&$2.40\times
10^{-2}$&$1.54\times 10^{-2}$&$1.05\times 10^{-2}$&$4.62\times 10^{-3}$\\
\hline
\end{tabular}
\begin{center}\caption{Detection probabilities
$P(m,a,z_s)$ \label{tab:prob}}\end{center}
\end{table*}
The results of applying Eq. (\ref{eq:pave}) for the three sets of
assumptions and two values of $z_s$ are given in Table \ref{tab:prob},
and Fig. \ref{fig:pmaplane} shows the probability distribution over
\begin{figure}[t]
\plotone{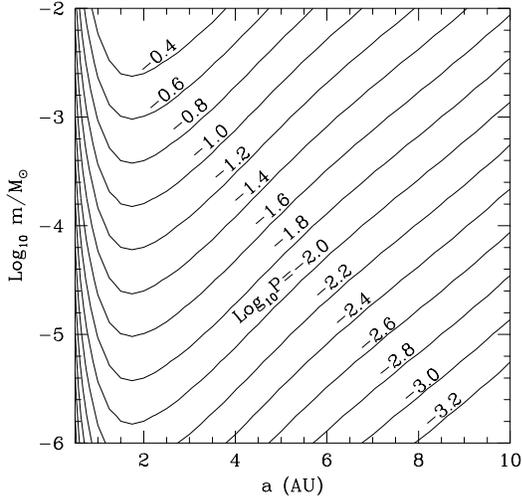}
\caption{Distribution over the $m,a$ plane of the probability of
detecting a planet $P(m,a,z_s)$ during an ongoing microlensing event
for the case where the average fractional distance to the lens
$z_s=0.8$ and where the planetary mass $m$ and semimajor axis $a$ are
assumed invariant over the lens mass distribution. \label{fig:pmaplane}}
\end{figure}
the $m,a$ plane for the model corresponding to the last column.
In this application of Eq. (\ref{eq:pave}), each planet type is
treated independently, so the probabilities include the perhaps
unlikely situation of detecting two planets around the same star.
Multiplying the number of assumed planets by these probabilities
yields the number of each of the planets that are detectable. We shall
assume that the fraction of the detectable small planets (Earths and
Venuses) that are actually characterized is the same as the 58\% averaged
maximum coverage, but that all of the detectable Jupiters are detected with the
probability curve in Fig. \ref{fig:probobs} now representing the average
fraction of the more than one day long light curve perturbation that
is followed with the high time resolution photometry. Saturn mass
planets will also have perturbation time scales comparable to a day
and all will be assumed detected, but Uranus perturbations will have
time scales comparable to the time the GC can be observed on an
average night at a single southern observatory, so perhaps 75\% of the
detectable Uranuses may be found. These assumptions are reflected in
the number of detected planets listed in Table \ref{tab:planets}.   
\begin{table*}[t]
\centering
\begin{tabular}{|c|c|c|c|c|c|c|}\hline
&\multicolumn{2}{|c|}{$m_i/M,\,a_i\sim$const.}&\multicolumn{2}{|c|}{$m_i/M,\,a_i/R_{\s E}\sim$const.}&\multicolumn{2}{|c|}{$m_i,\,a_i\sim$const.}\\\hline
planet&$z_s=0.5$&$z_s=0.8$&$z_s=0.5$&$z_s=0.8$&$z_s=0.5$&$z_s=0.8$\\
\hline
Venus&1&2&0&0&2&3\\  
Earth(1)&2&3&0&1&5&8\\
Earth(2.5)&3&3&2&4&6&5\\
Uranus(5)&10&6&22&17&14&9\\
Uranus(10)&2&1&5&4&2&1\\
Jupiter&60&38&131&100&85&52\\
Saturn&6&3&18&12&8&3\\
\hline
\end{tabular}
\begin{center}\caption{Number of planets detected.
\label{tab:planets}}\end{center}
\end{table*}

In the second of the three models in Table \ref{tab:prob}, where both the
$m/M$ and $a/R_{\s E}$ are invariant for each of the planets as the
lens mass varies, the probabilities simply reflect the scaled values
derived from Fig. \ref{fig:glprob}.  For example, the Jupiters in
this model always
correspond to a planet/lens mass ratio of 0.001 with a semimajor axis
that is always about 1.3 times the Einstein ring radius when $z=0.5$, where
its probability of detection is 0.175 for a lens of any mass. Hence,
the integral over the mass distribution (Eq. (\ref{eq:pave})) yields
just this probability in Table \ref{tab:prob}.  The detection
probability is reduced somewhat for $z=0.8$ in the table because the
semimajor axis is the same as for $z=0.5$ but the lensing zone has
moved in with the Einstein ring radius leaving the Jupiters to
the right of the peak in Fig. \ref{fig:glprob}.  Uranus at 5 AU for a
solar mass star occupies the same position as Jupiter, but its
probability of detection is scaled downwards by $\sqrt{m/M}$ to
the appropriate value of 0.039 for $z=0.5$ in the table.  The
probabilities for detecting Venuses and Earths nominally at 0.7 and 1
AU respectively (Nominally because these are the semimajor axes only
for $M=1$ in this model.) are negligibly
small because these planets are always too far inside the lensing zone in
this model, although one of the Earths is detected at 1 AU for $z_s=0.8$
and few more for both $z_s=0.5$ and 0.8 at 2.5 AU are detected.

The semimajor axes are fixed in the other two models, so the lensing
zone scans the planetary positions as the lens mass is reduced. For
the lower mass lenses this means that the Venuses and Earths at 0.7
and 1 AU move into the lensing zone so more are detected.  More
are detected when the masses are held constant because the planet/lens
mass ratio increases with decreasing lens mass. All of these very
close planets would be detected about stars less massive than the Sun.
The detection probabilities are higher for $z_s=0.8$ compared to
$z_s=0.5$ for the planets
that are close to their central stars and lower for those further
away because moving the lensing zone closer to the lens for the higher
value of $z_s$ moves the inner planets up the rising slope of the
probability curve inside the
lensing zone, while moving the outer planets further down the
decreasing curve outside the lensing zone. 

The entries into Tables \ref{tab:prob} and \ref{tab:planets} identify 
the planets detected, but the
data set obtained from the microlensing survey yields only the
planet-star mass ratio and the projected position of the planet in
units of the Einstein ring radius from the light curve alone. How much
information about the assumed distribution of planetary masses and
separations can be retrieved from the actual data set?  We begin to
answer this question by constructing that data set from the
distributions of the stellar masses and of the projected positions of the
planets.  Other information about the system that can be derived from 
different types of observations will be neglected for the time being.

We shall develop the sample data set only for the model corresponding to the 
last column in Tables \ref{tab:prob} and \ref{tab:planets}, where $z_s=0.8$ 
and $m_i$ and $a_i$ do not change with $M$. To determine the likely 
planet/star mass ratios for the detected events, we note from Eq. 
(\ref{eq:pave}) that the fraction of detected planets of given 
assumed mass and semimajor axis that orbit stars between masses $M=0.08$ and 
$M$ is given by 
\begin{equation}
f_{\s M}=\frac{\int_{0.08}^M{\cal F}(M^{\s\prime})P(m_i,M^{\s\prime}
,a_i,z_s)dM^{\s\prime}}
{\int_{0.08}^2{\cal F}(M^{\s\prime})P(m_i,M^{\s\prime},a_i,z_s)dM^{\s
\prime}}.\label{eq:fracofm}
\end{equation}
These cumulative fractions are shown in Fig. \ref{fig:fracofm} as a function  
of $M$. These fractions  depend only on the planet-lens 
\begin{figure}[t]
\plotone{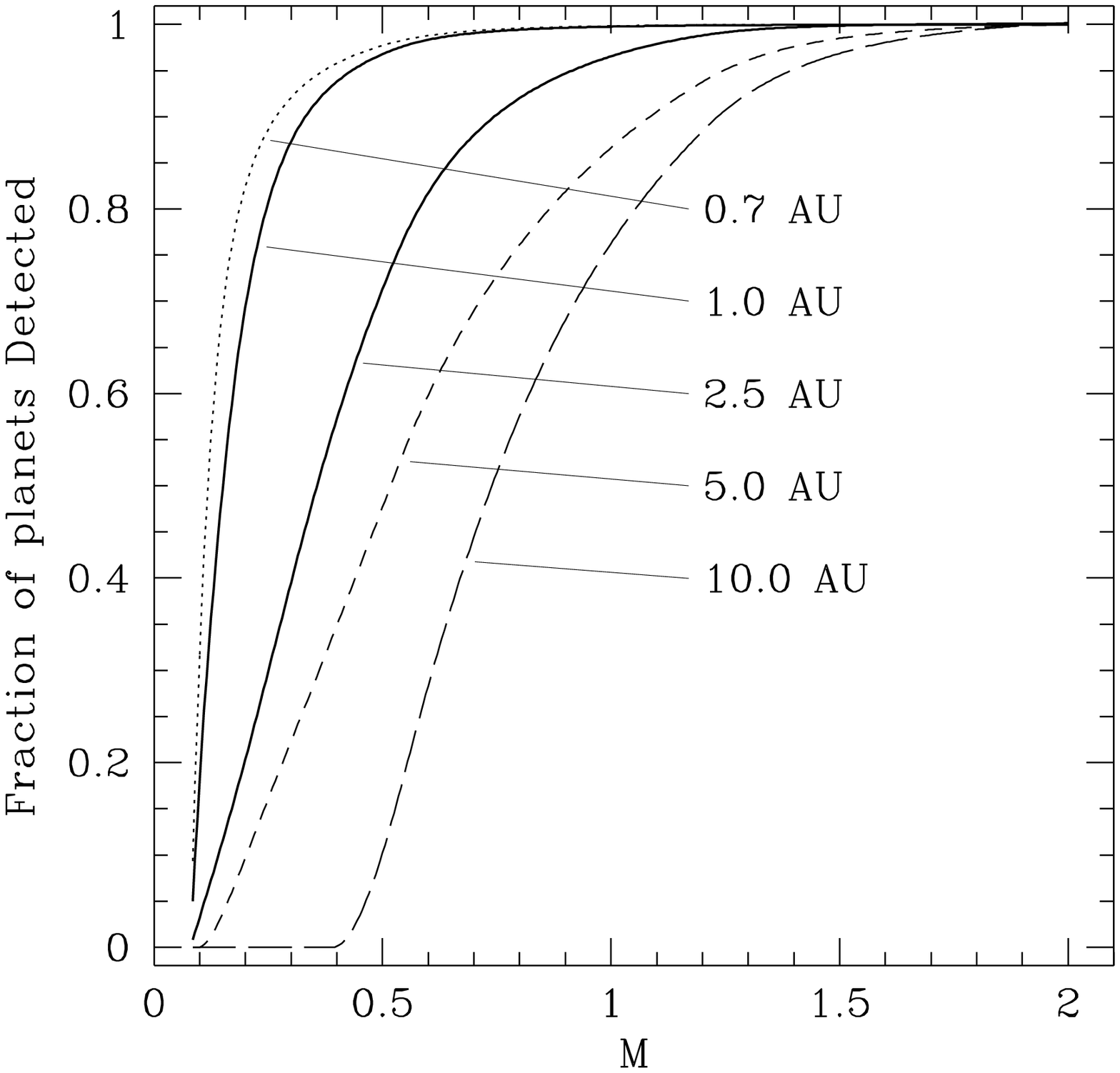}
\caption{Fraction of planets detected in the mass range 0.08 to $M$
for each semimajor axis in the model planetary systems. Close planets
are preferentially detected around the lower mass stars, whereas, the more
distant planets are detected over a wider range of
masses. \label{fig:fracofm}} 
\end{figure}
separation and not on the planetary masses, so the curves are labeled only 
with the semimajor axes. As mentioned 
earlier, the Venuses at 0.7 AU and the Earths at 1 AU are mostly detected  
around 
spectral class M stars, but the planets at 2.5 AU and beyond are distributed  
over a wider range of lens masses.  To assign the lens mass for each of the 
detected planets, we spread the detected masses over the distributions shown 
in Fig \ref{fig:fracofm} in proportion to the fraction detected in each range. 
For example, from Table \ref{tab:planets} the last column shows that about 
three Venuses would have been detected. From Fig. \ref{fig:fracofm}, about 1/3 
of the detected planets should be found about lenses in the mass range 
$0.08<M<0.10$, another third in $0.10<M<0.14$ and the remaining third in 
$M>0.14$. We can assign the probable lens masses for the Venus detections  
such that there is one within each range of masses corresponding to 1/3 of 
the probable detections. If we choose these to correspond to values of the 
cumulative fraction of detections near 1/6, 1/2, 5/6, the lens masses about 
which the Venuses are detected would be 0.09, 0.12, 0.20 respectively from
Fig. \ref{fig:fracofm}. We can systematically assign lens masses to $N$ 
detected planets with a particular semimajor axis by choosing those values 
of $M$ where the cumulative fraction of probable detections in Fig. 
\ref{fig:fracofm} is $1/(2N)+k/N,\; k=0,1,\cdots N-1$ as was done for Venus. 
The results of this exercise are shown in Table \ref{tab:data}. 
\begin{table*}
\centering
\begin{tabular}{cccc|cccc|cccc}
$M$&$m/M$&$r/a$&$r/R_{\s E}$&$M$&$m/M$&$r/a$&$r/R_{\s E}$&$M$&$m/M$&$r/a$&$r/R_{\s E}$\\
\hline
&&&&\multicolumn{4}{c|}{Venus (0.7 AU)}&&&&\\
0.09&$2.8\times 10^{-5}$&0.82&0.60&0.12&$2.1\times 10^{-5}$&0.95&0.59&0.20&$1.3\times 10^{-5}$&0.98&0.47\\
&&&&\multicolumn{4}{c|}{Earth (1.0 AU)}&&&&\\
0.09&$3.3\times 10^{-5}$&0.71&0.73&0.14&$2.1\times 10^{-5}$&0.91&0.75&
0.25&$1.2\times 10^{-5}$&0.98&0.61\\
0.10&$3.0\times 10^{-5}$&0.80&0.78&0.16&$1.9\times 10^{-5}$&0.94&0.73&
0.40&$7.5\times 10^{-6}$&0.99&0.48\\
0.12&$2.5\times 10^{-5}$&0.87&0.78&0.20&$1.5\times 10^{-5}$&0.96&0.66&&&&\\
&&&&\multicolumn{4}{c|}{Earth (2.5 AU)}&&&&\\
0.14&$2.1\times 10^{-5}$&0.55&1.14&0.36&$8.3\times 10^{-6}$&0.86&1.11&
0.74&$4.1\times 10^{-6}$&0.98&0.88\\
0.25&$1.2\times 10^{-5}$&0.75&1.17&0.49&$6.1\times 10^{-6}$&0.94&1.04&&&&\\
&&&&\multicolumn{4}{c|}{Jupiter (5.0 AU)}&&&&\\
0.12&$8.3\times 10^{-3}$&0.24&1.07&0.41&$2.4\times 10^{-3}$&0.61&1.47&
0.72&$1.4\times 10^{-3}$&0.87&1.59\\
0.14&$7.1\times 10^{-3}$&0.29&1.20&0.42&$2.4\times 10^{-3}$&0.63&1.50&
0.74&$1.4\times 10^{-3}$&0.89&1.60\\
0.16&$6.3\times 10^{-3}$&0.33&1.28&0.44&$2.3\times 10^{-3}$&0.64&1.49&
0.77&$1.3\times 10^{-3}$&0.90&1.59\\
0.18&$5.6\times 10^{-3}$&0.36&1.31&0.45&$2.2\times 10^{-3}$&0.65&1.50&
0.80&$1.3\times 10^{-3}$&0.91&1.58\\
0.19&$5.3\times 10^{-3}$&0.38&1.35&0.47&$2.1\times 10^{-3}$&0.67&1.51&
0.83&$1.2\times 10^{-3}$&0.92&1.56\\
0.21&$4.8\times 10^{-3}$&0.41&1.38&0.48&$2.1\times 10^{-3}$&0.68&1.52&
0.86&$1.2\times 10^{-3}$&0.93&1.55\\
0.22&$4.5\times 10^{-3}$&0.43&1.42&0.50&$2.0\times 10^{-3}$&0.70&1.53&
0.90&$1.1\times 10^{-3}$&0.94&1.53\\
0.24&$4.2\times 10^{-3}$&0.45&1.42&0.51&$2.0\times 10^{-3}$&0.71&1.54&
0.94&$1.1\times 10^{-3}$&0.95&1.52\\
0.25&$4.0\times 10^{-3}$&0.47&1.46&0.52&$1.0\times 10^{-3}$&0.73&1.57&
0.98&$1.0\times 10^{-3}$&0.96&1.50\\
0.27&$3.7\times 10^{-3}$&0.48&1.43&0.54&$1.9\times 10^{-3}$&0.74&1.56&
1.02&$9.8\times 10^{-4}$&0.97&1.49\\
0.28&$3.6\times 10^{-3}$&0.50&1.46&0.56&$1.8\times 10^{-3}$&0.76&1.57&
1.07&$9.3\times 10^{-4}$&0.98&1.47\\
0.30&$3.3\times 10^{-3}$&0.52&1.47&0.57&$1.8\times 10^{-3}$&0.77&1.58&
1.12&$8.9\times 10^{-4}$&0.99&1.45\\
0.31&$3.2\times 10^{-3}$&0.53&1.47&0.59&$1.7\times 10^{-3}$&0.79&1.59&
1.18&$8.5\times 10^{-4}$&0.99&1.41\\
0.33&$3.0\times 10^{-3}$&0.54&1.46&0.61&$1.6\times 10^{-3}$&0.80&1.59&
1.25&$8.0\times 10^{-4}$&0.99&1.37\\
0.34&$2.9\times 10^{-3}$&0.56&1.49&0.63&$1.6\times 10^{-3}$&0.81&1.58&
1.36&$7.4\times 10^{-4}$&0.99&1.31\\
0.36&$2.8\times 10^{-3}$&0.57&1.47&0.65&$1.5\times 10^{-3}$&0.83&1.59&
1.59&$6.3\times 10^{-4}$&1.00&1.23\\
0.38&$2.6\times 10^{-3}$&0.59&1.48&0.67&$1.5\times 10^{-3}$&0.84&1.59&&&&\\
0.39&$2.6\times 10^{-3}$&0.60&1.49&0.69&$1.4\times 10^{-3}$&0.86&1.60&&&&\\
&&&&\multicolumn{4}{c|}{Uranus (5.0 AU)}&&&&\\
0.17&$2.9\times 10^{-4}$&0.34&1.28&0.43&$1.2\times 10^{-4}$&0.64&1.51&
0.74&$6.8\times 10^{-5}$&0.89&1.60\\
0.25&$2.0\times 10^{-4}$&0.47&1.46&0.52&$9.6\times 10^{-5}$&0.72&1.55&
0.93&$5.4\times 10^{-5}$&0.95&1.52\\
0.34&$1.5\times 10^{-4}$&0.56&1.49&0.61&$8.2\times 10^{-5}$&0.81&1.61&
1.22&$4.1\times 10^{-5}$&0.98&1.37\\
&&&&\multicolumn{4}{c|}{Saturn (10.0 AU)}&&&&\\
0.54&$5.6\times 10^{-4}$&0.33&1.39&0.74&$4.1\times 10^{-4}$&0.60&2.16&
1.11&$2.7\times 10^{-4}$&0.93&2.73\\
&&&&\multicolumn{4}{c|}{Uranus (10.0 AU)}&&&&\\
&&&&0.74&$6.7\times 10^{-5}$&0.72&2.59&&&&\\
\end{tabular}
\begin{center}\caption{Probable distributions of planet/star mass
ratios and separations\label{tab:data}}\end{center}
\end{table*}

The determination of the distribution of normalized projected separations 
is more involved and not so well constrained as that for the planet/star mass 
ratios. An Earth with $a=2.5AU$ at its extreme projected separation would be
inside $R_{\s E}$ for a large star, but outside $R_{\s E}$ for a small star. 
But this Earth, uniformly distributed on a sphere of radius $a$, could be 
at any projected separation from the lens less than $a$ with a 
probability distribution given by $p(r)dr$ in Eq. ({\ref{eq:pdense}).
For a particular lens mass $M$ with a planet at a particular semimajor 
axis $a$, the probability of {\it detecting} the planet in a range $\Delta r$  
about $r<a$ is the probability of {\it finding} the planet in  
this range weighted by the probability of detecting the planet at 
that particular separation. The first probability peaks at $r=a$ and the second
peaks near the Einstein ring radius for the point sources considered here.
Although the probability distribution $P(m,a,M,z_s)$ obtained by scaling and
shifting the curve in Fig. \ref{fig:glprob} has been averaged over the
line of 
sight for a model distribution of stars, we can use it as an approximation to 
the second probability by replacing $a$ by $r<a$. It peaks near the Einstein 
ring radius and should give a reasonably good representation of the proper 
weighting of the projected distance probability.  The fraction $f_d$ of planets
of mass $m$, semimajor axis $a$ that would be detected in the range $0-r$ 
about an ensemble of lenses of mass $M$, average fractional distance $z_s$
is thus
\begin{equation}
f_d(r)=\frac{\int_0^r p(r^{\s\prime})P(m,r^{\s\prime},M,z_s)dr^{\s\prime}}
{\int_0^a p(r^{\s\prime})P(m,r^{\s\prime},M,z_s)dr^{\s\prime}}.
\label{eq:fracsep}
\end{equation}

In Fig. \ref{fig:fracsep} $f_d(r)$ is shown for intermediate values of
the lens mass $M$ in the lens mass distribution assigned to each
planet type. 
\begin{figure}[t]
\plotone{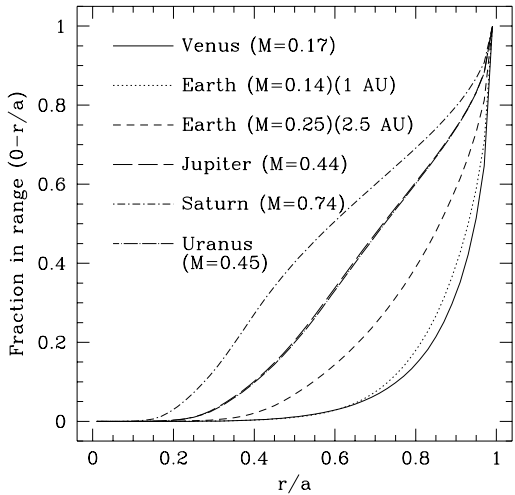}
\caption{Fraction of planets detected in range 0 to $r/a$ for each
planetary type in the model systems. Curves for each planet type
correspond to a lens mass near the mean value of the distribution of
lens masses about which that planet type were found in Table
\ref{tab:data}. They are the basis for the assigned distribution in projected
separations of planets. \label{fig:fracsep}}
\end{figure}
Three Venus mass planets are detected in our model data set about lenses with
masses 0.09, 0.12 and 0.20$M_\odot$ with corresponding Einstein ring radii of
0.97, 1.12, 1.44 AU respectively calculated at the average distance
$z_s=0.8$. The curves for the extreme values of $M$ for Venus in Fig. 
\ref{fig:fracsep} almost coincide with that for the intermediate value
shown for $M=0.12$. We therefore distribute the
distances of the detected Venuses according to this curve (like the  
distributions of lens masses) at values of $r/a=0.82,\,0.95,\,0.98$ 
corresponding to $f_d(r)\approx 0.16,\,0.5,\,0.83$ respectively.  We
assign the closer planets to the smaller lenses since the closer
proximity of the Einstein ring radii increases their probability of
detection.  The values of r/a and $r/R_{\s E}$ are given in Table 
\ref{tab:data}.  

The other planet types are treated in the same way with the
separations being distributed according to $f_d(r)$ for a lens mass 
giving a curve approximately midway between the curves for the extreme
values of the lens mass in the distribution. Here again the smallest  
separations are
associated with the smallest lenses with the values given in Table
\ref{tab:data}. As the semimajor axis gets small compared to the
Einstein ring radius the 
distribution is dominated by the probability of the projected
separation of the planet which peaks at $r=a$.  
The effect of the higher probability of detection at the Einstein ring
radius becomes important as the projected separation approaches this
radius and beyond. Hence, we find that $f_d(r)$ for all the planet
types moves toward the Venus curve when the lens mass is large.  For
the Earths at 1 AU at the largest lens mass $M=0.40$, about which one
of these closer Earths was detected, $f_d(r)$ almost coincides with
the Venus curve. For the Earths at 1 AU, the values of $r/a$ and
$r/R_{\s E}$ (at $z_s=0.8$) in Table \ref{tab:data} are distributed
according to the curve for $M=0.14$. Similarly, the Earths at 2.5 AU
are assigned separations according to the distribution for $M=0.25$, 
Jupiters according to the distribution for $M=0.44$, Saturns according
to the distribution for $M=0.74$, and Uranuses at 5 AU
according to the distribution for $M=0.45$.  The single Uranus that
was detected at 10 AU is placed at $r/a=0.72$ corresponding to $f_d(r)=0.5$
for $M=0.74$. This procedure gives a reasonably probable distribution
of separations, since the larger fraction detected at smaller
distances for smaller lenses is accounted for by assigning the closer
planets to the smaller lens masses.

The values of $m/M$ and $r/R_{\s E}$ in Table \ref{tab:data} 
comprise the information about the detected planets that can be learned from
the light curve.  Although additional information can be obtained by 
further observations of a different type, we start by seeing what can be 
deduced about the planetary systems from this information alone.

\section{Interpretation and discussion}

One of the first things that one notices about the data set in Table
\ref{tab:data} is that the mass ratios for the detected planets are 
reasonably good signatures for the actual masses of the planets in
spite of the varying lens masses. This results from the fact that the
lens masses only span an order of magnitude whereas the planetary
masses span three orders of magnitude.  We would thus conclude that
statistically  $m/M\sim 10^{-5}$ corresponds to terrestrial mass
planets, $m/M\sim 10^{-4}$ to Uranus mass planets and $m/M\sim 10^{-3}$
to Jupiter mass planets, where the terms terrestrial, Uranus and
Jupiter are meant to constrain the masses only to about an order
of magnitude from the least to the most massive in each
designation.  This grouping is evident in Fig. \ref{fig:mvsdist} where we
\begin{figure}[t]
\plotone{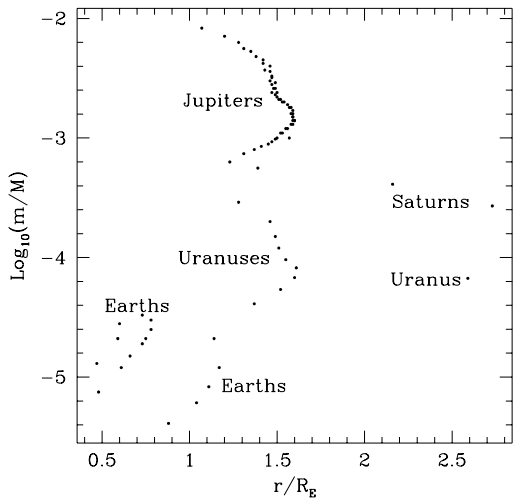}
\caption{Planet/lens mass ratios as a function of the projected
separation for the constructed data set in Table \ref{tab:data}. 
 \label{fig:mvsdist}} 
\end{figure}
have plotted the values of $m/M$ {\it vs} $r/R_{\s E}$. We also notice
that the lower mass planets tend to fall inside the Einstein ring
radii of their respective lenses, whereas the higher mass planets tend
to fall outside. Thus, this solar system characteristic of
our original assumptions is recovered. The concentration of the lower
mass planets near $r/R_{\s E}\approx 1$ are the Earths at 2.5 AU, and
we could infer this approximate location for these planets from the
fact that the probability of detection is highest when the planets have
semimajor axes near the Einstein ring radius and that radius is near
2.5 AU for the dominant number of small mass stars in the
distribution. The concentration of the medium and high mass planets
near $r/R_{\s E}=1.5$ is also striking, but this reflects the fact
that these planets have semimajor axes that are larger than the
Einstein rings of nearly all likely lenses, but have a greater
probability of being detected if their projected separations are
closer to the Einstein ring radius.  Two Saturns and the single
detected Uranus at 10 AU comprise the three points with $r/R_{\s E}>2$.

The facts that all the sources will not be 8 kpc from the Sun and that
the masses of real planets will not be limited to a few discrete
values would introduce additional scatter into the data set displayed
in Fig. \ref{fig:mvsdist}. The groups would no longer be evident,
although the general trend of the distribution of points would still
prevail if other planetary systems were to have small planets close to
their stars and larger planets further away. The very close large
planets around 51 Peg and 55 Cnc within .05 to .1 AU could not be
detected, but the very large planet about 47 UMa at 2.1 AU, if
detected, would yield a point in Fig. \ref{fig:mvsdist} most likely
between values of $r/R_{\s E}$ between 0.4 and 0.5 and $m/M>2.5\times
10^{-3}$ ($m\sin{i}=2.5\times 10^{-3}M_\odot$). If many points were
found in this region of
Fig. \ref{fig:mvsdist}, it would imply the common existence of large
planets considerably closer to their central stars than those in the
solar system.  

The detection of sixteen planets with planet/lens mass ratios near
$10^{-5}$ would imply that such planets are relatively common given
the low probability of their detection.  However, care must be
exercised in inferring the fraction of lenses having planets of
various masses from the number of planets detected and the
probabilities of detection, since the latter are dependent on the
assumptions about the planetary distributions and the lens mass
function. For example, with the same mass function, we detected
respectively 8, 5 and 16 terrestrial type planets for $z_s=0.8$ for
the three models, implying that our estimate of the fraction of lenses
having terrestrial mass planets could not be constrained by less than
a factor of 2.  Even though a factor of 2 may not sound too bad, the
small number of terrestrial planet detections makes such statistics
suspect. In any case the model dependence of the
interpretation is just as severe as that in deriving the expectations
from the search.  All the terrestrial planets are detected about stars
considerably less massive than
the Sun except higher mass lenses can be inferred for the terrestrial
planets detected with $r/R_{\s E}$ near 1.

The determination of the fraction of lenses having Jupiters is
similarly uncertain by at least a factor of 2. However, if nearly all of
these would correspond to values of $r/R_{\s E}>1$ as in our one data
set, we could infer the Jupiters to be relatively far away from their
stars consistent with formation near the ice condensation point in the
pre-planetary nebula.  This consistency with the current paradigm of
planet formation in the solar system and the inferred non dominance of
giant planet migration (Ward and Hourigan, 1989; Lin, {\it et al.}
1996; Ward, 1996)  in planetary system
histories allows us the further inference of the existence of several
terrestrial type planets inside the Jupiter orbit, even though the
latter were not directly detected.  The massive planets close to their
primaries can be detected by radial velocity and astrometric
techniques on the relatively short time scales of the orbit
periods as already demonstrated. Statistics for massive planets in
Jupiter-like orbits would be delayed in such surveys, but would start
accumulating immediately in an intensive microlensing search. An
occasional detection of a low mass planet would support the inference
that where there are Jupiters far out, terrestrial planets are in close. 

The detection of the nine Uranus mass planets just outside the Einstein Ring
radii in Table \ref {tab:data} and Fig. \ref{fig:mvsdist} would imply that
the Jupiters are not there for these systems. A further
inference could be that a relatively larger number terrestrial type
planets may be inside 5 AU for those particular stars than exist in our own
solar system (Wetherill, 1994).  The ratio of the number of
Uranuses  to the
number of Jupiters in such a data set would test the implication of
current theories that Jupiters at distances where they could form in the
nebula may be even more rare than we assumed at the outset. Radial
velocity and astrometric surveys can determine in a relatively short
time if many Jupiters have migrated to or are otherwise positioned
close to their stars.

We see that several conclusions can be drawn from the statistics of the
planet/star mass ratios and projected separations obtainable from
the microlensing light curves alone.  The fraction
of the stars that have planets of various masses in their lensing
zones can be estimated.  However, the model dependence of the interpretation
of these data limits the accuracy of the estimates to no better than a
factor of 2.  The planet/star mass ratios inferred from the data are
reasonably good indicators of the actual planetary masses because of
the limited range of likely stellar masses. A rough dependence of the 
planet-star separation on the planetary mass may be inferred from a
plot of $r/R_{\s E}$ {\it vs} $m/M$ (Fig. \ref{fig:mvsdist}).  The
number of Uranuses detected where Jupiters might be expected
constrains the success rate for the Jupiter-forming process. 

Few of the terrestrial type planets that are detected will
be close enough to their stars for habitability.  Recent developments
indicate that a way to infer the frequency of such planets may be the
distribution of separations of the massive planets from their stars.
About 1 in 30 stars in radial velicity surveys to date seem to have
giant planetary mass objects close to their primaries (W. Cochran, private 
communication 1996). There are several theoretical arguments showing how
Jupiters could be removed  from their region of formation and, in some
cases, account for the close planets. These include
torques from spiral density waves in a persistent nebula generated by 
fully formed planets (Lin {\it et al}, 1996), by planetary billiards
among giant planets formed too close together for stabilty (Rasio and
Ford, 1996, Weidenschilling and Marzari, 1996) or by rapid migration
of the solid body accretional cores toward the star before acquisition
of the gaseous envelopes (Ward and Hourigan, 1989; Ward, 1996).  

The
observations and the theories that attempt to account for them
imply that most of the material that would have
gone into terrestrial type planets could have been eliminated through
interactions with the giant planets or their cores that invaded the
terrestrial planet territory. Therefore, the greatest contribution that
a microlensing search for planets may make may be to determine where the
giant or subgiant planets are relative to their central stars. 
The absence of many Jupiters or Uranuses detected outside the Einstein
ring radii of their central stars may indicate that terrestrial-type
planets in habitable zones are infrequent. On the other hand, finding most
of a reasonably large number of massive planets relatively far from their
stars would indicate that the scattering or migration mechanisms did
not dominate planetary system histories and may give a better
inference of the frequency of occurence of
terrestrial planets than that provided by the meagre statistics of
the latter's direct detection.

We have limited the discussion so far to
just what may be learned from the statistics of the microlensing light
curves, since that is the only information that would be
systematically obtained in the microlensing search described above.
More information can be obtained about an individual system with
additional observations.  If the lens can be spectrally
classified, the actual masses of the planets of such
lenses can be obtained.  Although spectral classification allows an
estimate of the distance $D_{\s OL}$ to the lens from the known
luminosity and estimates of interstellar extinction, the additional
unknown $D_{\s OS}$ in $R_{\s E}$ limits the accuracy with which the
actual separation between lens and planet can be determined. 
An isolated star with visual magnitude
$m_v=24$ can be spectrally classified with a Keck telescope (R.M. Rich,
private communication, 1996). In the crowded field toward the galactic
center we should at least be able to classify a star with $m_v=22$,
which corresponds to a K5V main sequence dwarf at the galactic
center. The time after the event that one must wait in order to separate
the light of the lens from that of the source depends on the rate of
development of interferometric techniques, but we would know the
masses of those planetary companions of lenses earlier than K5V.  The
mass of a K5V star is about $0.69 M_\odot$ (Allen, 1973), so the
remaining lenses of later spectral type have a masses ranging over a
factor of 8.  

This is not that great an advantage since only about
25\% of the lenses in Table \ref{tab:data} have $M>0.69$. Still,
masses of nearly all of the detected planets 
would be known to within about a factor of 3 (an order of magnitude
from the least to the highest mass estimate) from a
microlensing search, and the projected separations in terms of the
Einstein ring radii would yield estimates of the actual
separations. In some circumstances observations using effects of differential
limb darkening of finite sized sources 
(Gould and Welch, 1996) or of limb brightened H-K lines of Calcium (Loeb and
Sassalov, 1995) as a small planetary Einstein ring traversed the source
might be used to gain additional information about the lens system.
However, discussion of the practical aspects of incorporating such
observations into an intensive microlensing survey should await
inclusion of the effects of finite sized sources in the detection of
small planets in an analysis like that above.  Inclusion of the finite
source size effects in a development similar to the above must await
the accumulation of an array of detection probability curves like
those of Bennett and Rhie (1996) covering a range of parameters.  A
possible approach to small planet detection with finite sized sources
is outlined in the next section.

\section{Finite sized sources}

Allowing for the finite size of the source changes the
procedure used to determine the planet/lens mass ratio for small planets, 
since the time scale 
of the planetary perturbation is now extended beyond the transit time
of a point source across the Einstein ring of the planet by the finite
angular size of the source star.  The maximum amplitude of the
perturbation of the light curve is also reduced since a small planet
can amplify only a fraction of an extended image.  The 
determination of the probability of a planetary detection (Bennett 
and Rhie, 1996) during an event is much different from that developed 
analytically by Gould and Loeb (1992). The Bennett 
and Rhie technique will be described below, and an example calculation of 
averaging a Bennett and Rhie detection probability over the stellar mass 
function establishes a procedure for determining the overall probability of 
detecting small mass planets. Even though this more correctly determined 
detection probability for small mass planets has little to do with that 
determined by scaling the Gould and Loeb probability curve for point sources, 
their remarkably close agreement for this one example implies that our 
conclusions above for the overall probability of detecting small mass planets 
may not be in error by more than a factor of 2. 

As pointed out above, the finite size of the source star
must be taken into account when the Einstein ring angular radius of the planet
becomes comparable with the angular size of the source star.
The angular size of a star at 8 kpc is $\theta_*=6\times
10^{-7}R_*/R_\odot$ arcsec, whereas the Einstein ring radius of a planet for
a source at 8 kpc is $1\times 10^{-3}\sqrt{m(1-z)/(M_\odot z)}=2\times
10^{-5}$ arcsec for a Jupiter at 6 kpc and $1\times 10^{-6}$ arcsec for an 
Earth at 6 kpc. So only for the largest super giants will the perturbation by a
Jupiter mass planet show the effects of the finite size of the source,
but essentially every perturbation by an Earth mass planet will be
greatly affected by the source size.  
Bennett and Rhie (1996) determine detection probabilities for
planet-star mass ratios of $10^{-5}$ and $10^{-4}$ for values of $z$
of 0.5 and 0.8 with source radii of $R_*=3$ and $13R_{\odot}$. We show
how these probabilities can be averaged over the mass function, but
that the probability determinations must be extended over the
parameter space before rigorous
overall probabilities of detection of Earth mass planets can be
found for the finite sized sources.

Fig. \ref{fig:brprob} shows the Bennett and Rhie (1996) probabilities
\begin{figure}[t]
\plotone{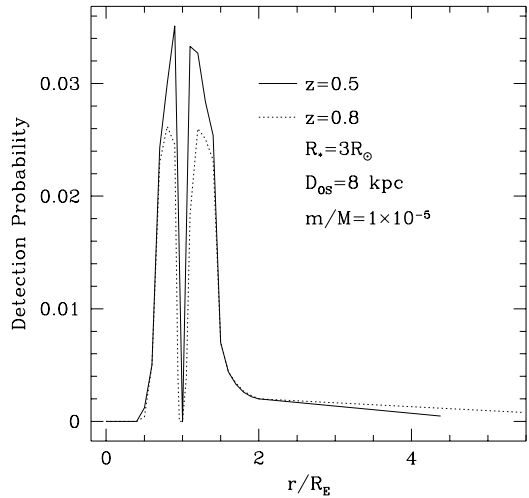}
\caption{Probability of detecting a planet as a function of the
projected lens-planet separation when $m/M=10^{-5}$ and the
source star at the center of the galaxy has a radius of
$3R_\odot$. (after Bennett and Rhie, 1996). The Bennett and Rhie
curves have been extended beyond $r=2R_{\s E}$ by determining the
probability of detecting a distant, essentially isolated planet given
that the planet's central star has acted as a lens. \label{fig:brprob}} 
\end{figure}
of detecting a planet with $m/M=10^{-5}$ (an Earth mass
when $M=0.3$) for the two lens distances as a function of
the planet's projected separation from its central star when the
source star at the center of the galaxy has a radius of
$3R_\odot$. This radius is a little larger than the peak in the distribution 
of radii of main 
sequence turn-off stars in the galactic bulge (Loeb and Sasselov, 1996).
The probabilities are determined by
numerically calculating many light curves for a set of impact
parameters of the source relative to the lens spanning the Einstein
ring of the lens and for all orientations of the source trajectory
across the ring relative to the lens-planet line. The fraction of the light
curves that displayed a perturbation more than 4\% of the unperturbed
amplification for more than 1/400 of the time scale of the entire event
gives the probability of detection. The effect of the
finite sized source is shown dramatically by the strong dip in the
detection probability when the projected separation of the planet is
near the Einstein ring radius. A point source yields a maximum
detection probability of detection when the planet is very near the
Einstein ring radius. This behavior for the finite sized source
follows because the image is spread out to its maximum size when it is
near the Einstein ring radius and the planet can focus only a much
smaller fraction of the image light---failing to reach threshold detection. 
It is of interest to note that the maximum probability of detection
for $z=0.5$ is approximately double the maximum scaled probability
for the point source case of Gould and Loeb (1992) (0.034 {\it vs.}
0.017 for $m/M=10^{-5}$), although the latter would be increased
somewhat if 4\% instead of 5\% were used as the detection criterion.
The Bennett and Rhie detection probability falls
somewhat for $z=0.8$ because the Einstein ring
radius of the planet is reduced at the larger value of $z$ so a
smaller fraction of the source image is amplified by the planet for
similar geometries.

Below we shall average these
probabilities over the mass function of the lenses for planet orbital
semimajor axes 0.7, 1.0, 2.5 and 5 AU respectively.  The largest
semimajor axis corresponds to a separation of $5.5R{\s E}$ for
$M=0.08$ and $z=0.8$. We have therefore extended the Bennett and
Rhie curves from $r=2R_{\s E}$ to $5.5R{\s E}$ by estimating the
probabilities of detecting the distant planet (Appendix B).

The functional dependence of the probability curves is
$P=P(m,M,R_*,z,r)$, where $r$ is the projected separation of the planet
from the lens in the lens plane and $R_*$ is the radius of the source
(assumed to be at 8 kpc distance).
The probability of detecting a planet of mass $m$ with semimajor axis
$a$ during an event with source radius $R_*$, lens of mass $M$ located
at fractional distance $z$ is the probability of finding the planet at
projected distance between $r$ and $r+dr$ (Eq. (\ref{eq:pdense})) times
$P(m,M,R_*,z,r)$ integrated from 0 to $a$ or 
\begin{equation}
P(m,M,R_*,z,a)=\int_0^a\frac{P(m,M,R_*,z,r)rdr}{\sqrt{1-\frac{r^2}{a^2}}},
\label{eq:brrtoa} 
\end{equation}
where all distances are determined in terms of the Einstein ring
radius. So in general the integral is a function of $M$ for given
$m,R_*,z,a$. If the probability curves are left in the broken
line form shown in Fig. \ref{fig:brprob}, each linear segment substituted into
Eq. (\ref{eq:brrtoa}) is analytically integrable and the singularity
in the integrand at $r=a$ is removed. Similar to the Gould and Loeb
probabilities, the probability of detecting a planet of mass $m$ with
semimajor axis $a$ as a companion of a lens at $z$ in an event with
source radius $R_*$ averaged over the event mass distribution is
\begin{equation}
P(m,R_*,z,a)=\int_{0.08}^2 {\cal F}(M)P(m,M,R_*,z,a)dM,\label{eq:brprobma}
\end{equation}

The curves in Fig. \ref{fig:brprob} correspond to a particular mass
ratio, so we can only evaluate the averaged probability in
Eq. (\ref{eq:brprobma}) for the case where the planet/star mass ratio
is fixed at this particular value as the lens mass varies. 
These probabilities are shown in
Table \ref{tab:bravep} along with those for point sources.
\begin{table*}[t]
\centering
\begin{tabular}{|l|l|l|l|l|}\hline
&\multicolumn{2}{c|}{$P(m,R_*,z,a)$ (Eq. (\ref{eq:brprobma}))}&\multicolumn{2}{c|}{$P(m,a,z_a)A$ (Eq. (\ref{eq:pave}))}\\
\hline
$a$ AU&z=0.5&z=0.8&z=0.5&z=0.8\\
\hline
0.7&0.000155&0.000992&0.00133&0.00289\\
1.0&0.00239&0.00405&0.00421&0.00685\\
2.5&0.0158&0.0121&0.0140&0.0133\\
5.0&0.0105&0.00592&0.00797&0.00510\\
\hline
\end{tabular}
\begin{center}\caption{Probabilities of detecting a planet with 
$m/M=10^{-5}$ and
$R_*=3R_\odot$ with probabilities derived for point sources with $m/M$
fixed at $10^{-5}$ included for comparison \label{tab:bravep}}
\end{center}
\end{table*}
With the exception of $a=0.7AU$, the probabilities for point and
finite sources are remarkably close.  However, we
cannot draw any inferences from these comparisons except perhaps to
think that the probabilities for detecting Earth mass planets
when point sources are assumed are closer to the real values than we
might have expected.  The finite source probabilities must be averaged
over the distribution of lenses along the line of sight and over the
distribution of sizes of those sources that will be monitored in any
microlensing survey.  This requires the generation of curves like
those in Fig. \ref{fig:brprob} for a sufficient number of source sizes
and lens distances for reasonable interpolation over the
distributions. In addition, if we were to consider the model where the
mass of the planet is held constant instead of the mass ratio,
detection probability curves
must be generated for various mass ratios.  Hence, more accurate
estimates of the detection of Earth mass planets through microlensing
must await generation of this array of Bennett and Rhie type
probability curves or, as may be unlikely, a clever scheme of
avoiding  the lengthy
calculations can be contrived. On the other hand, the comparison of
the probabilities in Table \ref{tab:bravep} with those from the point
source calculations, may indicate that our estimates in Tables
\ref{tab:prob} and \ref{tab:planets} for the Earth mass planets are
correct to within a factor of 2. 

For the point source, the planet only perturbs the single lens light curve 
significantly when it is within a planetary Einstein ring radius of one of 
the unperturbed 
image positions in the lens plane. If the angular size of this Einstein ring 
is small compared with that of the source star, it will be small compared 
with the size of the image in the lens plane. Nothing much happens as the 
planetary Einstein ring approaches the edge of a large image, since it can 
amplify only a small 
fraction of the image light.  That fraction of course depends on the angular 
size of the image (or angular size of the source star) 
compared to that of the planetary Einstein ring, so bigger 
perturbations occur for smaller sources.  Generally, a significant 
perturbation occurs only as the planetary Einstein ring moves well into the 
finite image, which occurs, at least for a planet outside the lens Einstein 
ring radius, as the caustic curve in the source plane moves across the face 
of the source star. The picture is more complicated if the planet is
inside the
lens Einstein ring radius since there are two caustics giving positive 
amplification separated by a deep negative trough, If the finite sized source 
covers both caustic and negative trough, the resulting cancellation can keep 
the total amplification undetectably small (Bennett and Rhie, 1996). If the 
source angular diameter is less than the separation of the two caustics a 
negative perturbation of the light curve can be detected for the inside 
planet. 

In most cases the perturbation of the light curve will have
more structure than that shown by the example of
Fig. \ref{fig:planltcv},  since regions of positive amplification are
always interleaved with regions of negative amplification. Examples of
such light curves for several planet/lens mass ratios, several
geometries and several source sizes are shown by Bennett and Rhie
(1996) and by Wambsganss (1996). The perturbation light curve
structure is most pronounced when the source size is smaller than the
caustic in the source plane, but persists if the source is not too
large, where a typical pattern is a negative perturbation
surrounded by smaller positive perturbations for a planet inside the
Einstein ring of the lensing star and a positive perturbation
surrounded by smaller negative perturbations for a planet outside the
Einstein ring (Bennett and Rhie, 1996).  

The structure of the light curve perturbation complicates the
simple picture of determining the planet/ lens mass ratio from the
ratio of the perturbation time scale and the event time scale, but at
the same time supplies sufficient constraints for a more precise
determination of that mass ratio.  The normalized impact parameter of the
lens-source encounter is known from the maximum of the light curve, and
the angle at which the source encounters the lens-planet line and the
location of the unperturbed images of the source are known
from the position of the perturbation on the light curve. The projected
position of the planet must be near one of the unperturbed images in
order effect the perturbation. The selection of the correct image is
usually very easy with a well sampled light curve as noted above
(Gaudi and Gould, 1996). The planet may pass either side of the image
and create the same maximum perturbation, but with different
symmetries in the wings of the light curve that are nearly always
discernable for passage near the image inside the Einstein ring (minor
image) and usually discernable for passage near the outside image (major
image) with 1\% photometry and good coverage (Gaudi and Gould, 1996).
If the ratio of the source angular radius to the angular radius of the
planetary Einstein ring is not too large, a condition satisfied by
Earth mass planets when the source is a turn-off main sequence star at
the center of the galaxy ($R_*\sim 2\,{\rm to}\,3R_{\s\odot}$), the
remaining uncertainty in the projected position of the planet is
negligibly small. As the angular size of the source is known
spectroscopically from the apparent brightness and an estimate of
interstellar absorption, the planet/star mass ratio, as the remaining free
parameter, can be adjusted to match the light curve and thereby
constrain its value quite well.  A model of limb darkening and slight
adjustments in the source trajectory through the lens-planet system
may be necessary to refine the determination. 

About 25\% of the events toward the galactic center will involve 
so-called clump giants ($R_*\sim 13 R_{\s\odot}$) as sources (Alcock,
{\it et al} 1996). Gaudi and Gould (1996) call attention to a
potentially serious uncertainty in the planet/star mass ratio by
demonstrating a set of similar, low amplitude 
perturbation light curves involving the major image with the same
maximum change in amplification and the same full-width-half-maximum
of the main part of the perturbation but corresponding to an order of
magnitude range of planet/star mass ratios.  The uncertainties are
reduced to acceptability with good coverage of both wings of the light
curves with 1\% photometry. However, the practical difficulties in
getting  such coverage routinely means that the uncertainties would
remain for many of these events.  One means of reducing the 
uncertainties in the mass ratio to acceptable values offered by Gaudi
and Gould is to distinguish the curves with simultaneous photometry in
the infrared and optical. Giant stars are less limb darkened in the
infrared than in the optical, and the distinct color effects of the
event can select the proper mass ratio.  On the other hand, the
planetary perturbations are more difficult to detect in the first place with
giant sources because of the lower amplitude of the perturbations. In
most of the cases were the uncertainty would be severe, the observing
strategy used by Bennett and Rhie (1996) would have failed to detect
the planet (D. Bennett, private communication 1996). This would
eliminate most of the cases where the uncertainty in $m/M$ was
excessive from the data set.  But even with those that remain, there
are ways to determine both $m/M$ and the projected separation to
sufficient accuracy (Gaudi and Gould, 1996).

So the effect of finite source size will not hinder the determination of
the planet/lens mass ratio or the normalized projected separation of
the planet from the lens.  It thus makes sense to pursue the exercise outlined 
above to determine the overall probability of detection of small mass planets 
when the finite source size is accounted for.  In the meantime, we can take 
some assurance from the comparison of the detection probabilities for finite 
and point sources in Table \ref{tab:bravep}, that our estimates for the 
detection of small planets may be within a factor of 2 of those that will be 
derived for the finite sources.

We end by cautioning the reader to
keep in mind the severe assumption dependence of the expectations for
planet discoveries in any microlensing search. (Table \ref{tab:planets}
gives approximately 56, 138, or 81 planets detected for the three sets
of assumptions assuming that no planets are found about members of
binary star systems.) But he or she should
also keep in mind that the detection scheme is robust with simple,
ground based technology, and that if they are there, the planets will
be detected with useful bounds placed on their masses and planet-star
separations with a data set based on the light curves alone. Finally,
although microlensing is the only ground based scheme that is sensitive to
Earth-mass planets, finding most of a large number of giant or
semi-giant planets
beyond the Einstein ring radii of their stars may be a better
indicator of the frequency of occurrence of terrestrial planets than
the few of the latter that are directly detected. 
\appendix
\begin{center}Appendices\end{center}
\section{Analytic Representation of the Gould and
Loeb (1992) detection probability curve}
\begin{displaymath}
x=\frac{0.25a}{\sqrt{M}}\sqrt{\frac{0.25}{z(1-z)}}
\end{displaymath}
\begin{eqnarray*}
P&=&0,\qquad 0\leq x\leq 0.1\\
\\
&=&0.001763066\left[e^{7.6653342*(x-0.1)}-1.0\right],\\
&&0.1\leq x\leq 0.564\\
\\
&=&0.4733727x-0.2069822,\qquad 0.564\leq x\leq 0.733\\
\\
&=&-84.9463918x^4+ 286.1509291x^3-361.2286639x^2\\
&&+202.6160685x-42.4667434,\quad 0.733\leq x\leq 0.9\\
\\
&=&0.0475524d0*x+.1204028,\quad 0.9\leq x\leq 1.043\\
\\
&=&-1.3858063x^4+6.0121987x^3-9.8567924x^2\\
&&+7.2772186x-1.8790527,\quad 1.043\leq x\leq 1.4\\
\\
&=&-0.1806818x+0.4164545,\quad 1.4\leq x\leq 1.664\\
\\
&=&0.02869257x^4-0.3053152x^3+1.23616560x^2\\
&&-2.2872802x+1.6857629, \quad 1.164\leq x\leq 3.0\\
\\
&=&-0.015d0*x+0.075,\qquad 3.0\leq x\leq 5.0\\
\\
&=&0,\qquad x\geq 5.0\\
\end{eqnarray*}
\section{Probability of detection of a planet that is
far from its central star}
Bennett and Rhie (1996) require an amplification of 1.58 before a
microlensing event is to be followed with high time resolution
photometry in search of a planet. This requires that the source have
an impact parameter relative to the lens of less than 0.7637 according
to Eq. (\ref{eq:amp}). A distant planet ($a>>R_{\s E}$) will act as a
solitary lens for most of its probable projected distances from the
lens. The planetary Einstein ring radius $R_{\s E_p}$ is small compared
to the projected radius of the source star, which requires that the
planet actually traverse the face of the source star for significant
amplification.  (This compares with its having to traverse the image of
the source star for significant amplification when it is close to its
central star.) The maximum amplification is found by aligning  the
planet with the center of the source, multiplying Eq. (\ref{eq:amp})
by an element of area of the source (normalized by the Einstein ring
radius) integrating over the total area of the source and dividing by
the area. This yields 
\begin{displaymath}
A_{max}=\sqrt{1+\frac{4\theta_{R_{\s Ep}}^2}{\theta_*^2}},
\end{displaymath}
where $\theta_{\s R_{\s Ep}}$ is the angle subtended by the Einstein
ring radius of the planet and $\theta_*$ is that subtended by the
source radius $R_*$.  For the mass $10^{-5}\times 0.08M_\odot$ with
$R_*=3R_\odot$, $A_{max}=1.60$ for $z=0.5$, and $A_{max}=1.125$ for
$z=0.8$. In both cases the planet would be detected by the 4\%
criterion, but only after the event because of the $A=1.58$ threshold
before a planet is sought.  These maximum amplifications also show why it
is necessary for the planet to be on the disk of the source star
before detection is possible. 

The planet must be within the swath that is $2\times .7637R_{\s E}$ wide
and $2a$ long centered on the lens star to have a chance of being
intercepted by the source during the event. However, since the distant
planet can be detected only after the event, we only consider the
probability $P_s$ that the planet is in one half of this swath. 
\begin{eqnarray*}
P_s&=&\frac{1}{2}\int_0^R\frac{r\,dr}{a^2\sqrt{1-r^2/a^2}}\\
&&+\int_R^a
dr\int_{-{\rm sin}^{-1}R/r}^{{\rm sin}^{-1}R/r}\frac{r}{a^2\sqrt
{1-r^2/a^2}}\frac{d\theta}{2\pi}
\end{eqnarray*}
$P_s=0.0494$ for $z=0.8$ where the swath is $2\times 0.915\times 5\,
({\rm AU})^2$, and $P_s=0.0635$ for $z=0.5$ where the swath is $2\times
1.14\times 5\, ({\rm AU})^2$. 

For $z=0.8$, $3R_\odot$ projected onto the lens plane and divided by
$R_{\s E}$ for $M=0.08$ is $r_*=0.0122$. If the planet is within the swath
area it is approximately uniformly distributed in the direction
perpendicular to the long axis of the swath. The probability that the
source will encounter the planet is then $r_*/0.7637=0.016$ and the
probability of detection is this probability times the probability
that the planet is within the swath or 0.00079.  This establishes the
point (5.46,0.00079) to which the Bennett and Rhie graph for $z=0.8$ is
extended in Fig. \ref{fig:brprob}.  For $z=0.5$, $r_*=0.00614$, the
probability of interception is 0.00803 and probability of detection is
0.00051. So the Bennett and Rhie graph for $z=0.5$ is extended to the
point (4.37, 0.00051) in Fig. \ref{fig:brprob}.

\begin{center}Acknowledgements\end{center}
\acknowledgements
I am very grateful to E. Agol, D. Bennett, O. Blaes, A. Gould,
A. Loeb and S. Rhie for reading a draft of the manuscript,
pointing out some errors and offering suggestions for improving the
presentation.  I especially thank A. Gould and A. Loeb for their
support over the last year, while I studied the art of the
microlensing detection of planets. Special thanks are due David Tytler
who motivated a concentrated effort toward microlensing planet
detection. This work was supported by the NASA
PG\&G program under grant NAGW 2061.

\parindent=0pt
\parskip=5pt
\begin{center}References\end{center}

Alard, C., S. Mao, and J. Guibert (1995) Object DUO 2 - a new binary lens 
candidate, {\it Astron. Astrophys.} {\bf 300}, L17-L20.

Alcock, C., R.A. Allsman, T.S. Axelrod, D.P. Bennett, K.H. Cook, K.C. Freeman,
K. Griest, S.L. Marshall, S. Perlmutter, B.A. Peterson, M.R. Pratt, P.J. Quinn,
A.W. Rodgers, C.W. Stubbs, and W. Sutherland (1995) Probable gravitational 
microlensing toward the galactic bulge, {\it Astrophys. J.} {\bf 445}, 133-139.

Alcock, C., R.A. Allsman, T.S. Axelrod, D.P. Bennett, K.H. Cook, K.C. Freeman,
K. Griest, J. Guern, M.J. Lehner, S.L. Marshall, H.-S Park,
S. Perlmutter, B.A. Peterson, M.R. Pratt, P.J. Quinn,
A.W. Rodgers, C.W. Stubbs, and W. Sutherland (1996) The Macho project:
45 candidate microlensing events from the first year of galactic bulge
data, {\it Astrophys. J.} In Press.

Allen, C.W. (1973) {\it Astrophysical Quantities}, Althone Press, U. of 
London, p 209.

Basu, S. and N.C. Rana (1992) Multiplicity-corrected mass function of 
main-sequence stars in the solar neighborhood, {\it Astrophys. J.} {\bf 393},
373-384.

Beichman, C. A. (Ed.) (1996) {\it A Road Map for the Exploration of 
Neighboring Planetary Systems}, Jet Propulsion Lab Publication 92-22,
Pasadena, CA.

Bennett, D.P. and S.H. Rhie (1996) Detecting Earth-mass planets with 
gravitational microlensing, {\it Astrophys. J. Let.} In Press. 

Boss, A.P. (1996) Giants and dwarfs meet in the middle, {\it Nature} {\bf 379},
397-398.

Butler, R.P. and G.W. Marcy (1996) A planet orbiting 47 UMa, {\it Astrophs. 
J. Let.} In Press. 

Chang, K., and  S. Refsdal (1979) Flux variations of QSO 0957+561A,B
and image splitting by stars near the light path, {\it Nature} {\bf 282}, 561.

Feller, W. (1957) {\it An Introduction to Probability Theory and its
Applications}, John Wiley and Sons, New York.

Gatewood, G. (1996) Lalande 21185, {\it Bull. Amer. Astron. Soc.} {\bf 28}, 
885.

Gauli, B.S. and A. Gould (1996) Planet parameters in microlensing
events, {\it Astrophys. J.}, Submitted.

Gould, A. (1996) Microlensing and the stellar mass function, {\it Proc. Astron.
Soc. Pac.} In press.

Gould, A. and D.L. Welch (1996) Macho proper motions from optical/infradred 
photometry, {\it Astrophys. J.} {\bf 464}, 212-217.

Gould, A., and A. Loeb (1992) Discovering planetary systems through 
gravitational microlenses, {\it Astrophys. J.} {\bf 396}, 104-114.

Lin, D.N.C., P. Bodenheimer and D.C. Richardson (1996) Orbital migration of 
planetary companion of 51 Pegasi to its present location, {\it Nature} {\bf 
380}, 606-607.

Loeb, A. and D. Sassalov (1995) Removing Degeneracy of microlensing light 
curves through narrowband photometry of giants, {\it Astrophys. J. Let.} 
{\bf 449}, L33-L36.

Mao, S. and B. Paczy\'nski (1991) Gravitational microlensing by double stars 
and planetary systems, {\it Astrophys. J.} {\bf 374},L37-L40.

Marcy, G.W. and R. P. Butler (1996) A planetary companion to 70 Vir, 
{\it Astrophs. J. Let.} In Press.

Mayor, M. and D. Queloz (1995) A Jupiter-mass companion to a solar-type star, 
{\it Nature} {\bf 378}, 355-359.

Mazeh, T., M. Mayor and D.W. Latham (1996) Eccentricity vs. mass for low-mass
companions and planets, Submitted.

Rasio, F.A. and E.B. Ford (1996) Dynamical instabilities and the
formation of extrasolar planetary systems, {\it Science} {\bf 274}, 954-956.

Schneider, P., and A. Weiss (1986) The two-point-mass lens: detailed 
investigation of a special asymmetric gravitational lens, {\it Astron. 
Astrophys.} {\bf 164}, 237-259. 

Strom, K.M., J. Kepner, S.E. Strom (1995) The evolutionary status of the 
stellar population in the Rho Ophiuchi cloud core, {\it Astrophys. J.} {\bf 
438}, 813-829.

Tytler, D. (1995) "Exploration of Neighboring Planetary Systems (ExNPS):
Ground Based Element," WWW address: 
http://techinfo.jpl.nasa.gov/WWW/\\ExNPS/RoadMap.html. Also in 
{\it A Road Map for the Exploration of 
Neighboring Planetary Systems}, Jet Propulsion Lab Publication 92-22,
Pasadena, CA, 1996.

Ward, W.R. (1996) Survival of planetary systems, {\it Bull. Amer. 
Astron. Soc.} {\bf 28}, 1112.

Ward, W.R., and K. Hourigan (1989) Orbital migration of protoplanets - The inertial limit, {\it Astrophys. J.} {\bf 347}, 490-495.

Wambsganss, J. (1996) Discovering galactic planets by gravitational
microlensing: magnification patterns and light curves, {\it
Mon. Not. Roy. Astron. Soc.} In press.

Weidenschilling, S.J. and F. Marzari (1996) A possible origin of giant
planets found at small stellar distances, Submitted to {\it Nature}.

Wetherill, G. (1994) Possible consequences of the absence of ``Jupiters'' in 
planetary systems, In {\it Planetary Systems: Formation, Evolution and 
Detection} (B.F. Burke, J.H. Rahe, and E.E. Roettger, Eds.) pp 23-32. Kluwer, 
Dordrecht. 

Witt, H.J. (1990) Investigation of high amplification events in light curves
of gravitationally lensed quasars, {\it Astron. Astrophys} {\bf 236}, 311-322.

Zhao, H.S., D.N. Spergel and R.M. Rich (1995) Microlensing by the galactic bar,
{\it Astrophys. J.} {\bf 440}, L13-L16.

\end{document}